\documentclass{emulateapj}

\usepackage{amsmath}
\usepackage{amssymb}
\usepackage{latexsym}
\usepackage{textcomp}
\usepackage{graphicx}
\usepackage{subfigure}
\usepackage{lscape}
\usepackage{longtable}
\usepackage{natbib}
\newcommand{\angstrom}{${\rm \AA}$}

\newcommand{\loiii}{$L_{[{\rm O}\,\text{\tiny III}]}$}
\newcommand{\oiii}{[O\,{\tiny III}]}
\newcommand{\oii}{[O\,{\tiny II}]}
\newcommand{\heii}{He\,{\tiny II}}
\newcommand{\mgib}{Mg\,{\tiny I}\,$b$}
\newcommand{\neiii}{[Ne\,{\tiny III}]}

\slugcomment{Submitted 2009 April 24; accepted 2009 July 20}

\begin{document}
\title{Host Galaxies of Luminous Type 2 Quasars at $z \sim 0.5$\altaffilmark{1}}

\shorttitle{HOST GALAXIES OF LUMINOUS TYPE 2 QUASARS}
\shortauthors{LIU ET AL.}

\author{\sc Xin Liu\altaffilmark{2}, Nadia L. Zakamska\altaffilmark{3,4}, Jenny E. Greene\altaffilmark{2,5}, \\
Michael A. Strauss\altaffilmark{2}, Julian H.
Krolik\altaffilmark{6}, and Timothy M. Heckman\altaffilmark{6}}

\altaffiltext{1}{Based, in part, on observations obtained at the
Gemini Observatory, which is operated by the Association of
Universities for Research in Astronomy, Inc., under a cooperative
agreement with the NSF on behalf of the Gemini partnership: the
National Science Foundation (United States), the Science and
Technology Facilities Council (United Kingdom), the National
Research Council (Canada), CONICYT (Chile), the Australian
Research Council (Australia), Minist\'{e}rio da Ci\^{e}ncia e
Tecnologia (Brazil) and Ministerio de Ciencia, Tecnolog\'{i}a e
Innovaci\'{o}n Productiva  (Argentina).}

\altaffiltext{2}{Department of Astrophysical Sciences, Princeton
University, Peyton Hall -- Ivy Lane, Princeton, NJ 08544}

\altaffiltext{3}{Institute for Advanced Study, Einstein Dr.,
Princeton, NJ 08540}

\altaffiltext{4}{Spitzer Fellow, John N. Bahcall Fellow}

\altaffiltext{5}{Hubble Fellow, Princeton-Carnegie Fellow}

\altaffiltext{6}{Department of Physics and Astronomy, Johns
Hopkins University, Baltimore, MD 21218}

\begin{abstract}
We present deep Gemini GMOS optical spectroscopy of nine luminous
quasars at redshifts $z \sim 0.5$, drawn from the SDSS type 2
quasar sample. Our targets were selected to have high intrinsic
luminosities ($M_V < -26$ mag) as indicated by the \oiii\
$\lambda$5007 ${\rm \AA}$ emission-line luminosity (\loiii). Our
sample has a median black hole mass of $\sim 10^{8.8} M_{\odot}$
inferred assuming the local $M_{{\rm BH}}$-$\sigma_{\ast}$
relation and a median Eddington ratio of $\sim 0.7$, using stellar
velocity dispersions $\sigma_{\ast}$ measured from the G band. We
estimate the contamination of the stellar continuum from scattered
quasar light based on the strength of broad H$\beta$, and provide
an empirical calibration of the contamination as a function of
\loiii; the scattered light fraction is $\sim 30$\% of $L_{5100}$
for objects with \loiii\ $= 10^{9.5} L_{\odot}$. Population
synthesis indicates that young post-starburst populations ($< 0.1$
Gyr) are prevalent in luminous type 2 quasars, in addition to a
relatively old population ($> 1$ Gyr) which dominates the stellar
mass. Broad emission complexes around \heii\ $\lambda$4686 ${\rm
\AA}$ with luminosities up to $10^{8.3} L_{\odot}$ are
unambiguously detected in three out of the nine targets,
indicative of Wolf-Rayet populations. Population synthesis shows
that $\sim$5-Myr post-starburst populations contribute
substantially to the luminosities ($>50$\% of $L_{5100}$) of all
three objects with Wolf-Rayet detections. We find two objects with
double cores and four with close companions. Our results may
suggest that luminous type 2 quasars trace an early stage of
galaxy interaction, perhaps responsible for both the quasar and
the starburst activity.
\end{abstract}


\section{Introduction}\label{sec:intro}

Most, if not all, bulge-dominated galaxies harbor supermassive
black holes \citep[SMBH;][]{kormendy95,magorrian98}. Studying the
host galaxies of the most luminous quasars is essential for
understanding the coupled evolution of SMBHs and galaxies. The
linked growth has been strongly informed by the similar redshift
evolution of the space density of quasars
\citep[e.g.,][]{boyle98,hasinger05,richards06} and that of the
star formation rate (SFR) \citep[e.g.,][]{connolly97,chapman05},
and the correlation between black hole mass and bulge stellar
velocity dispersion (the $M_{{\rm BH}}$-$\sigma_{\ast}$ relation)
found in local inactive galaxies \citep{ferrarese00,gebhardt00}.

Simulations of gas-rich mergers \citep[e.g.,][]{dimatteo05}
ascribe SMBH-galaxy co-evolution to merge-induced starburst and
quasar activity that in turn outputs energy (``feedback'')
regulating further growth \citep{silk98,fabian99}. While these
models claim to be successful in interpreting multiple
observations \citep[such as the quasar luminosity function and the
distribution of Eddington ratios; e.g.,][]{hopkins06,shen09}, the
predicted brief phases of concurrent starburst and quasar
activity, in spite of being crucial for the growth of stellar
populations and SMBHs, are rarely observed.

There is growing evidence that low-luminosity AGN are not mainly
induced by major mergers \citep[e.g., based on close neighbor or
galaxy lopsidedness
studies:][]{fuentes88,derobertis98,pierce07,ellison08,li08,reichard09}.
For the most luminous quasars ($L > 10^{46}$ erg s$^{-1}$), on the
other hand, gas-rich mergers have long been implicated, as
supported by the close companions and tidal tails seen in early
HST studies of quasar host galaxies
\citep[e.g.,][]{bahcall97,kirhakos99}, but the merger scenario
remains controversial \citep[e.g.][]{floyd04,bennert08,urrutia08}.

In a study of $\gtrsim 20,000$ type 2 AGN with \loiii$/L_{\odot}
\sim 10^{5.0}$--$10^{8.5}$ (extinction corrected) at $z < 0.3$
selected from the Sloan Digital Sky Survey
\citep[SDSS;][]{york00}, \citet{kauffmann03} found that the host
galaxies of AGN with higher luminosities have younger mean stellar
ages than a control sample of inactive galaxies, while those of
AGN with lower luminosities have stellar ages similar to those of
normal early-type galaxies. At higher luminosities (\loiii~$>
10^{8.5}L_{\odot}$), it is unclear whether luminous quasars
preferentially reside in massive elliptical galaxies with little
recent star formation activity
\citep[e.g.,][]{mclure99,nolan01,dunlop03}, or if there is
considerable recent or on-going star-forming activity
\citep[e.g.,][]{boroson82,heckman97,canalizo00,jahnke04,letawe07}.

Studies of the stellar populations and interaction rates in the
most luminous quasars are significantly hampered by the high
contrast between nuclear and stellar light. Techniques to
circumvent this problem include taking off-nucleus spectra
\citep[e.g.][]{hutchings90,nolan01,watson08,wolf08} and modeling
and subtracting nuclear light in observed on-nucleus spectra
\citep[e.g.][]{magain05,letawe07,jahnke07}. Here we focus on
luminous obscured quasars, for which the obscuring material acts
as a natural coronagraph, allowing detailed study of the host
galaxy morphology and stellar populations.

Due to the redshift evolution and the shape of the quasar
luminosity function \citep[e.g.][]{boyle00,richards06},
low-redshift quasars comparable to luminous high-redshift quasars
are quite rare, and in addition the obscured ones are hard to
find. \citet{reyes08} have recently published a sample of $\sim$
100 type 2 quasars with \loiii$>10^{9.3}L_{\odot}$ (among a total
of $\sim$ 900 type 2 quasars with \loiii$>10^{8.3}L_{\odot}$).
This paper presents deep Gemini optical spectroscopy of a pilot
sample of nine luminous type 2 quasars with \loiii\, $> 10^{9.3}
L_{\odot}$ at redshift $z \sim 0.5$. We determine the relative
contribution of old and young stellar populations to probe the
starburst-quasar link. We measure stellar velocity dispersions,
allowing us to determine black hole masses assuming the $M_{{\rm
BH}}$-$\sigma_{\ast}$ relation \citep{tremaine02}. We combine
these measurements with estimates of quasar intrinsic luminosities
to obtain accretion rates.

The strict interpretation of AGN unification models
\citep[e.g.,][]{antonucci93} states that type 1 and type 2 quasars
are identical aside from our viewing angle. Here we test a
prediction of unification by quantifying the level of scattered
light from the accretion disk present in type 2 quasars, as seen
so dramatically in the polarization measurement of
\citet{zakamska05} and the pilot imaging survey of \citet[][see
also \citealt{cid04}]{zakamska06}. Of course, this scattered light
component is a substantial contaminant to our estimates of ongoing
star formation, and it is thus critical to quantify. We also
address the more fundamental issue of whether or not strict
unification holds; a substantial fraction of type 2 quasars may
well suffer from galaxy-scale obscuration associated with vigorous
star formation, making the type 2 population as a whole more
biased toward star-bursting populations
\citep[e.g.][]{martinez06,rigby06,lacy07}.

The paper is structured as follows. We discuss sample selection in
\S \ref{sec:selection}, and describe our Gemini observations and
data reduction in \S \ref{sec:obs}. Our data analysis method and
results are provided in \S \ref{sec:result}. We present
implications and discussion in \S \ref{sec:discuss}, and summarize
our main conclusions in \S \ref{sec:sum}. Throughout we use AB
magnitudes and assume a cosmology with $\Omega_m = 0.3$,
$\Omega_{\Lambda} = 0.7$, and $h = 0.7$.

\section{Sample Selection: the Most Luminous Subgroup of SDSS Type 2
Quasars}\label{sec:selection}

Our targets were drawn at the high-luminosity end of a parent
sample of $\sim 900$ optically-selected type 2 quasars in the SDSS
\citep{zakamska03,reyes08}. The luminosity of the \oiii\
$\lambda$5007 emission line, \loiii, is adopted as a proxy for the
intrinsic quasar luminosity. Arising from the narrow-line region,
the forbidden line \oiii\ $\lambda$5007 should be much less
affected by circum-nuclear obscuration; its luminosity is observed
to be correlated with the broad-band continuum luminosity in
unobscured quasars \citep[e.g.][]{reyes08}.

We need sufficient spectral resolution and high enough
signal-to-noise ratio (S/N) in the continuum to obtain robust
measurements of stellar velocity dispersions. The SDSS spectra
have adequate spectral resolution ($R \sim 2000$), but the
continuum S/N for most objects at the high-luminosity end is
inadequate, typically $<3$ pixel$^{-1}$. Based on \loiii\,
measured from the SDSS spectra, we selected a pilot sample of nine
objects with \loiii\, $> 10^{9.3} L_{\odot}$ (corresponding to
intrinsic luminosities $M_V < -26$ mag; see below) for deep
follow-up optical spectroscopy. In addition, all of our targets
were selected to be radio-quiet to avoid complications from radio
jets in interpreting the results, and to be at redshifts $z \sim
0.5$ for proper wavelength coverage. There is no additional
selection criterion other than the visibility of the target at the
observing time. In particular, we did not select on continuum
flux, so these objects are representative of the parent sample at
any fixed \loiii. The targets are listed in Table \ref{tab:obs} in
increasing RA order.

\section{Observation and Data Reduction}\label{sec:obs}

\begin{deluxetable*}{ccccccc}
\tabletypesize{\scriptsize}
\tablewidth{0pc}
\tablecaption{
Type II Quasars Observed with Gemini GMOS
\label{tab:obs}
}
\tablehead{
\colhead{~~~~~~~~~~~~Target Name~~~~~~~~~~~~} &
\colhead{~~~~~~Redshift~~~~~~} &
\colhead{~~~~~~R.A.~~~~~~} &
\colhead{~~~~~~Decl.~~~~~~} &
\colhead{~~~~~~$r$~~~~~~} &
\colhead{~~~~~~\loiii~~~~~~} &
\colhead{Exposure} \\
\colhead{(1)} &
\colhead{(2)} &
\colhead{(3)} &
\colhead{(4)} &
\colhead{(5)} &
\colhead{(6)} &
\colhead{(7)}
}
\startdata
 SDSSJ0056$+$0032\dotfill & $0.4840$ & 00 56 21.72 & $+$00 32 35.8 & 20.10$\pm$0.03 & 9.34$_{-0.00}^{+0.01}$  & 14400      \\
 SDSSJ0134$+$0014\dotfill & $0.5550$ & 01 34 16.34 & $+$00 14 13.6 & 20.72$\pm$0.04 & 9.63$_{-0.04}^{+0.00}$  & ~3600      \\
 SDSSJ0157$-$0053\dotfill & $0.4213$ & 01 57 16.92 & $-$00 53 04.8 & 20.42$\pm$0.03 & 9.25$_{-0.00}^{+0.00}$  & 14400      \\
 SDSSJ0210$-$1001\dotfill & $0.5398$ & 02 10 47.01 & $-$10 01 52.9 & 20.27$\pm$0.03 & 9.92$_{-0.01}^{+0.00}$  & ~3600      \\
 SDSSJ0319$-$0058\dotfill & $0.6261$ & 03 19 50.54 & $-$00 58 50.6 & 21.32$\pm$0.07 & 9.76$_{-0.20}^{+0.00}$  & ~3600      \\
 SDSSJ0801$+$4412\dotfill & $0.5560$ & 08 01 54.24 & $+$44 12 34.0 & 21.13$\pm$0.05 & 9.53$_{-0.00}^{+0.11}$  & ~3600      \\
 SDSSJ0823$+$3231\dotfill & $0.4332$ & 08 23 13.50 & $+$31 32 03.8 & 20.12$\pm$0.03 & 9.77$_{-0.00}^{+0.02}$  & ~3600      \\
 SDSSJ0943$+$3456\dotfill & $0.5293$ & 09 43 11.57 & $+$34 56 15.9 & 20.07$\pm$0.02 & 9.84$_{-0.00}^{+0.04}$  & ~5400      \\
 SDSSJ0950$+$0511\dotfill & $0.5231$ & 09 50 19.91 & $+$05 11 40.9 & 18.99$\pm$0.02 & 9.45$_{-0.00}^{+0.26}$  & ~3600      \\
\enddata
\tablecomments{ \\
Col.(1): Target name in the format ``SDSSJhhmm$\pm$ddmm''. \\
Col.(3),(4): J2000 coordinates. \\
Col.(5): $r$-band model magnitude and uncertainty from SDSS (uncorrected for Galactic extinction). \\
Col.(6): \oiii\ $\lambda$5007 emission-line luminosity in the
form of log$(L/L_{\odot})$ measured over the continuum-subtracted
GMOS spectra. The total uncertainty
is calculated as the convolution of the 1-$\sigma$ measurement
error and the systematic uncertainty estimated using the
difference between the GMOS and SDSS observations. A comparison
between SDSS and Gemini spectroscopic measurements is presented in
Figure \ref{fig:compare}. \\
Col.(7): Total exposure time in seconds.  }
\end{deluxetable*}

\subsection{Follow-up Observation with Gemini GMOS}\label{subsec:obs}

\begin{figure*}
    \centering
        \includegraphics[width=130mm]{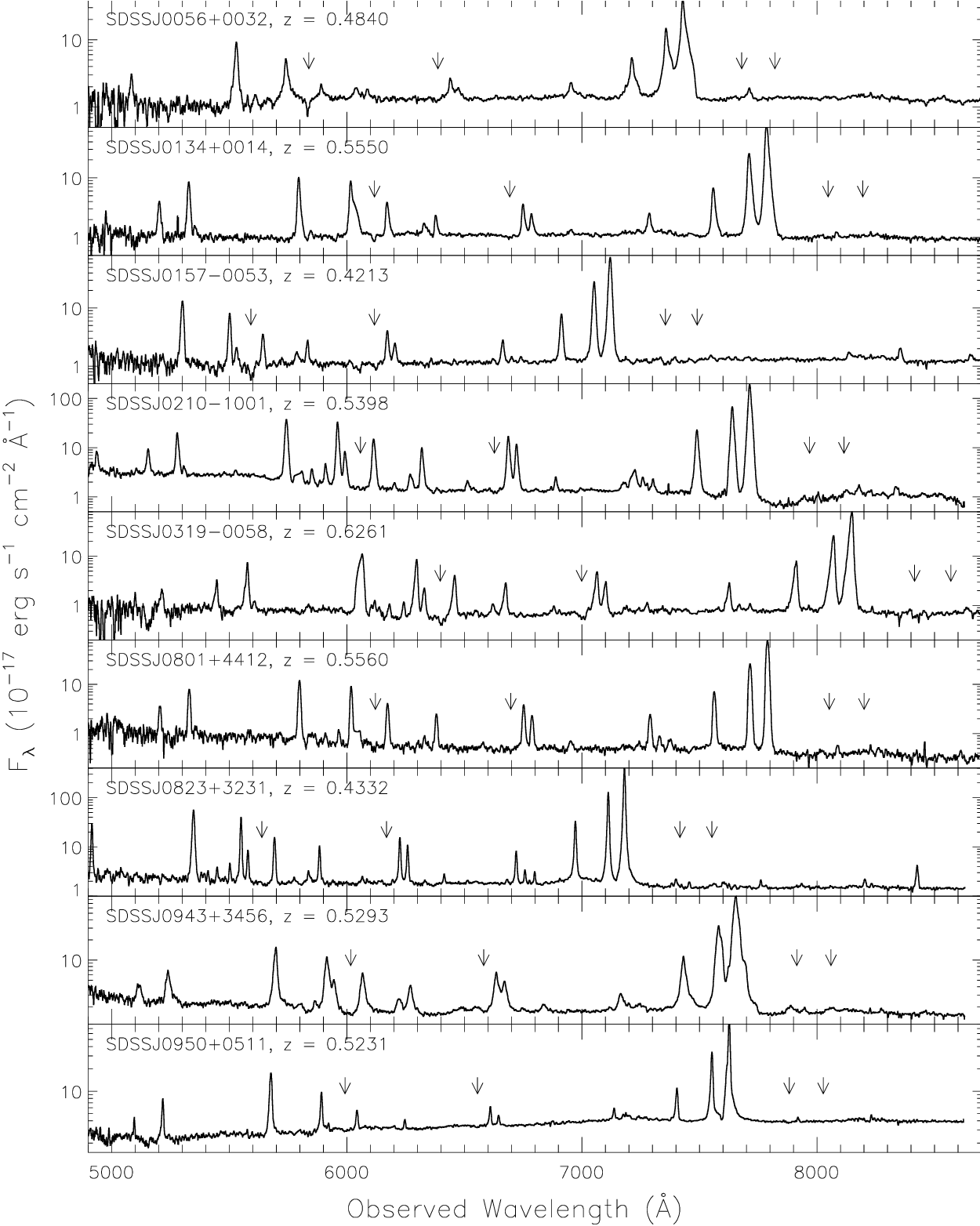}
    \caption
    {
    Flux-calibrated Gemini GMOS spectra (smoothed by a 6-pixel boxcar for display purpose)
    of the nine targets in this study.
    Downward arrows indicate the expected positions of stellar absorption features
    Ca K $\lambda3934$ \angstrom, G band $\lambda4304$ \angstrom, \mgib\ triplet $\lambda5175$ \angstrom,
    and Fe $\lambda5270$ \angstrom.
    Ca K and the G band are discernable in most cases. The \mgib\ triplet and Fe, on the other hand, show no clear detections
    except in SDSSJ0157$-$0053, most likely due to the contamination by adjacent emission lines from Fe and N
    ions and their intrinsically smaller equivalent widths.
    In almost all targets, \oiii\ $\lambda\lambda4959,5007$ \angstrom\ lines show significant
    asymmetry and deviation from Gaussian.
    }
    \label{fig:spec_obscor}
\end{figure*}

Optical long-slit spectra were obtained for the nine targets using
the Gemini Multi-Object Spectrograph (GMOS) on Gemini-North on 12
nights between August 2006 and January 2007 (program ID:
GN-2006B-0493). The seeing was variable, ranging from 0.4 to 1.1
arcsec. The slit-width used was 0.5 arcsec, corresponding to $\sim
3$ kpc (physical) at redshifts $z \sim 0.5$ in the assumed
cosmology. The spatial sampling of the detector was 0.29 arcsec
pixel$^{-1}$. The total exposure time for each target ranged from
1 to 4 hours (with 30 minutes per exposure) depending on weather
conditions, and is listed in Table \ref{tab:obs}. The median S/N
per rest-frame 1.0 \angstrom\, was 22, 32, and 43 over the
spectral ranges 3900--3960, 4250--4320, and 5150--5350 \angstrom.
These S/N \angstrom$^{-1}$ achieved in GMOS spectra of our targets
are at least 10 times larger than those of their SDSS spectra. The
slit was centered on the quasar itself and oriented to cover as
many objects in the field as possible to observe potential
companions and/or extended-emission line regions.

The R400-G5305 grating was adopted with a dispersion of $\sim$
0.45 \angstrom\, pixel$^{-1}$ and a spectral resolution of
$R$$\sim$ 1900, spanning an observed wavelength range of
5000--8000 \angstrom. At the redshifts of our targets, this
corresponds to rest-frame 3300--5300 \angstrom, a range covering
the prominent signatures of both old and young stellar populations
(see \S \ref{subsec:pop}). An A0 white dwarf (G191B2B) was
observed as a flux standard. We also observed a spectroscopic
standard star (a K-giant) to calibrate the instrumental resolution
(see \S \ref{subsec:sigma} for our calibration approach).

\subsection{Data Reduction}\label{subsec:redux}

The reduction of the 2D spectra was performed in IRAF using the
Gemini GMOS package. The main steps to reduce each science
exposure before flux calibration include bias subtraction, flat
fielding, interpolation across the chip gaps, cleaning for cosmic
rays and bad pixels, wavelength calibration using arc exposures
taken right before and after observing each target, sky
subtraction, and extraction of the one-dimensional spectrum. The
spectra were extracted using a 5 arcsec aperture, which was
determined from the extension of the point spread function (PSF).

The science spectra were then flux calibrated using the
photometric standard star and corrected for atmospheric extinction
using the curve appropriate for the Gemini observatory with IDL
routines as described in \citet{matheson08}. The spectra were
shifted to the heliocentric frame. Different exposures of the same
target were calibrated separately before being co-added since they
had different air masses and observing epochs. The resulting 1-D
spectra of the nine targets are displayed in Figure
\ref{fig:spec_obscor}.

\begin{figure}
    \centering
        \includegraphics[width=42mm]{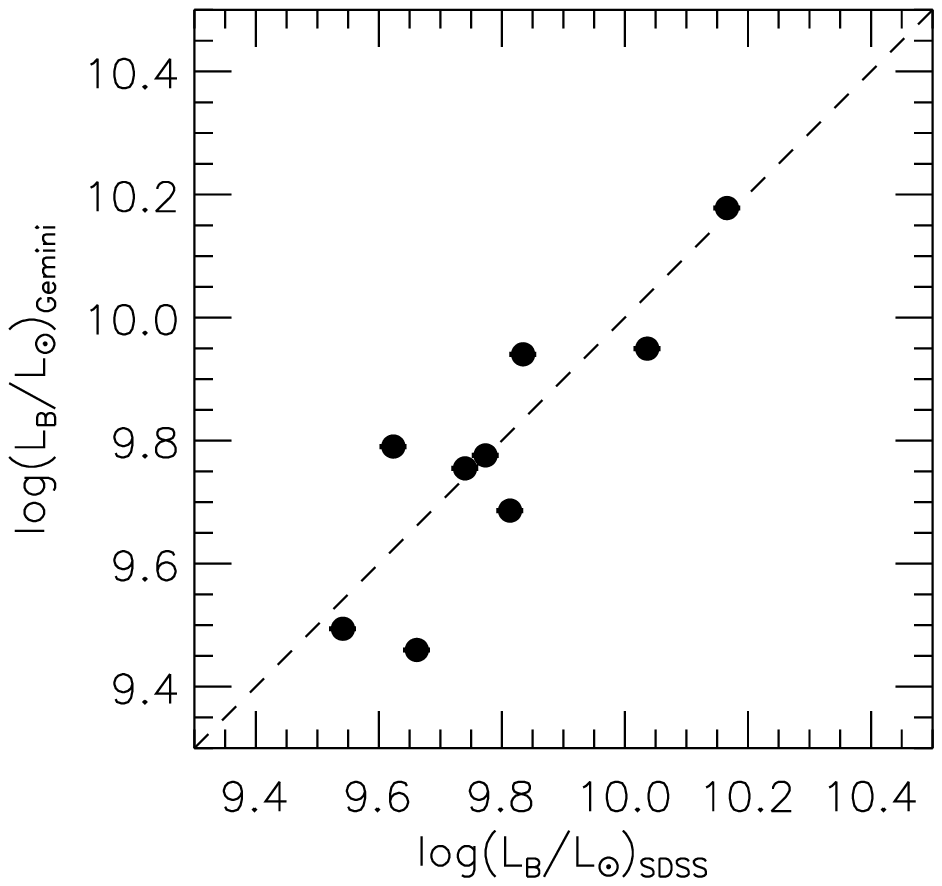}
        \includegraphics[width=42mm]{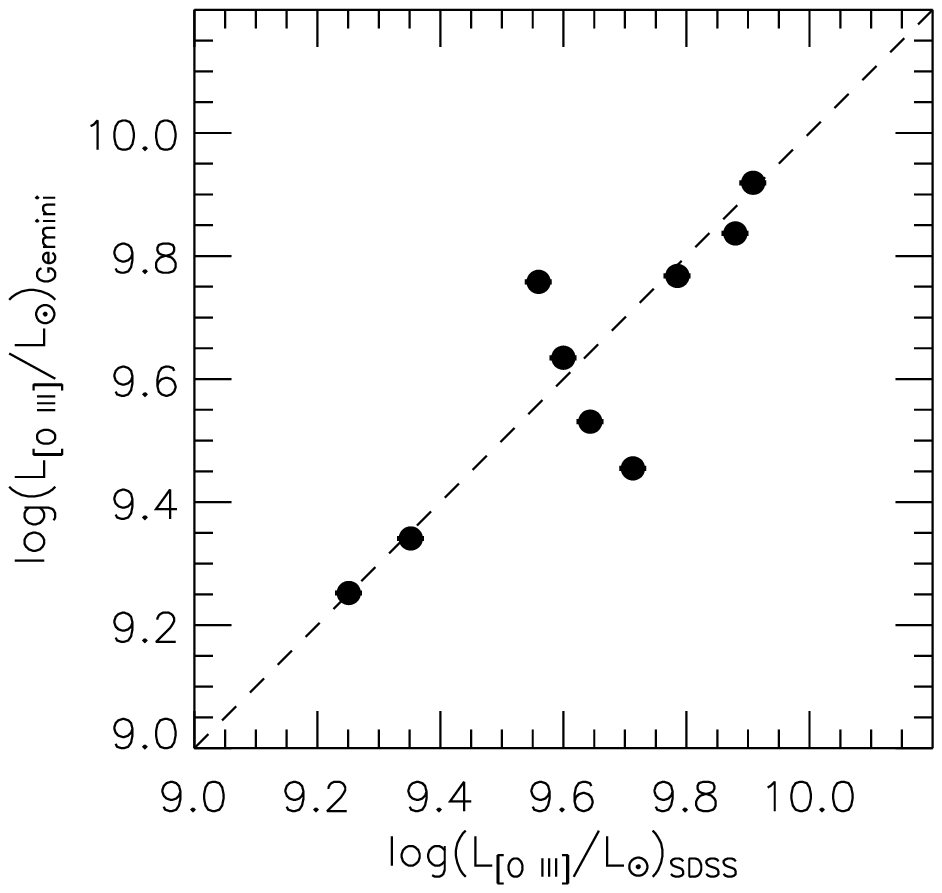}
    \caption{
    Continuum and emission-line luminosities measured from GMOS and SDSS spectroscopic observations.
    The SDSS observations are made from 3-arcsec-diameter fibers, whereas the GMOS long-slit spectra were extracted
    using a 5-arcsec aperture with a slit width of 0.5 arcsec. The $B$-band luminosity $L_B$ is
    calculated as that emitted between rest-frame 3980--4920 ${\rm \AA}$ \citep{zakamska03}.
    The emission-line \oiii\ $\lambda$5007 luminosity \loiii\, is obtained from nonparametric
    fits using the model profile of \oiii\ lines \citep{reyes08}.
    The symbol size corresponds to $\pm5$\% calibration errors.
    The continuum and emission-line measurements from the two observations agree,
    given the systematics due to the difference between long-slit
    and fiber coverage and seeing slit losses.
    }
    \label{fig:compare}
\end{figure}

We check the spectrophotometry of our targets against their SDSS
spectra, of which the spectrophotometry for point sources is
better than 5\% \citep{dr6}. The general agreement between the two
observations in both continuum and emission-line luminosities is
illustrated in Figure \ref{fig:compare}. While on average the two
measurements agree, for several objects they differ from each
other by more than 5\%. We have checked that the continua of the
GMOS and SDSS spectra of each of our targets are in reasonable
agreement, whereas the strongest emission lines \oiii\
$\lambda\lambda$4959,5007 ${\rm \AA}$ differ both in shape and in
amplitude for several objects. Since we know the \oiii\ is
spatially more extended, the line-shape differences presumably
reflect the different spatial coverage of the SDSS fiber
(3\arcsec\ diameter) and the Gemini long-slit. Indeed, 2-D spectra
show that the spatial extent of the strong emission lines is
usually larger than that of the continuum. We conclude that the
differences between the GMOS and SDSS measurements, if any, are
most likely due to the difference in the long-slit and fiber
coverage and slit losses due to seeing variations. We take these
differences as the estimate of systematic uncertainties. The
systematic error added in quadrature to the measurement error is
taken as the total uncertainty for the luminosity measurement.

\section{Data Analysis and Results}\label{sec:result}

In this section, we present our data analysis methods and results.
We measure stellar velocity dispersions in \S \ref{subsec:sigma},
estimate scattered light from the strength of broad H$\beta$ in \S
\ref{subsec:scatteredlight}, analyze stellar populations in \S
\ref{subsec:pop}, and present double cores, companions and/or
extended emission line regions covered by our long-slit
observations in \S \ref{subsec:companion}.

\begin{deluxetable*}{cccccccc}
\tabletypesize{\scriptsize}
\tablewidth{0pc}
\tablecaption
{
Stellar Velocity Dispersions, Black Hole Masses, and Eddington Ratios
\label{tab:sigmafit}
}
\tablehead{
\colhead{~~~~~~~~~~~~Target Name~~~~~~~~~~~~} &
\colhead{~~~$\sigma_{\ast}^{{\rm G}}$~~~} &
\colhead{~~~$\sigma_{\ast}^{{\rm Ca}}$~~~} &
\colhead{~~~$\sigma_{\ast}^{{\rm Mg}}$~~~} &
\colhead{~~~~~~$M_{{\rm BH}}$~~~~~~} &
\colhead{$M_{{\rm BH}}^{{\rm vir}}$} &
\colhead{~~~~~~$L_{{\rm Bol}}$~~~~~~} &
\colhead{~~~$L_{{\rm Bol}}/L_{{\rm Edd}}$~~~} \\
\colhead{(1)} &
\colhead{(2)} &
\colhead{(3)} &
\colhead{(4)} &
\colhead{(5)} &
\colhead{(6)} &
\colhead{(7)} &
\colhead{(8)} }
\startdata
 SDSSJ0056$+$0032\dotfill  &  119$\pm$16  & 104$\pm$13   & \nodata    & 7.2$\pm$0.2 & \nodata     &  12.8$_{-0.5}^{+0.5}$ & 12$_{-9}^{+26}$        \\
 SDSSJ0134$+$0014\dotfill  &  180$\pm$21  & 157$\pm$33   & \nodata    & 7.9$\pm$0.2 & \nodata     &  13.1$_{-0.5}^{+0.5}$ & 4.4$_{-3.4}^{+9.8}$    \\
 SDSSJ0157$-$0053\dotfill  &  212$\pm$15  & 229$\pm$27   & 182$\pm$9  & 8.2$\pm$0.1 & \nodata     &  12.7$_{-0.5}^{+0.5}$ & 0.90$_{-0.65}^{+2.0}$  \\
 SDSSJ0210$-$1001\dotfill  &  346$\pm$43  & \nodata      & \nodata    & 9.1$\pm$0.2 & 9.2$\pm$0.4 &  13.3$_{-0.5}^{+0.5}$ & 0.50$_{-0.39}^{+1.1}$  \\
 SDSSJ0319$-$0058\dotfill  &  286$\pm$15  & 299$\pm$35   & \nodata    & 8.8$\pm$0.1 & \nodata     &  13.2$_{-0.6}^{+0.5}$ & 0.85$_{-0.66}^{+1.9}$  \\
 SDSSJ0801$+$4412\dotfill  &  250$\pm$43  & \nodata      & \nodata    & 8.5$\pm$0.3 & 9.1$\pm$0.4 &  12.9$_{-0.5}^{+0.6}$ & 0.73$_{-0.62}^{+2.3}$  \\
 SDSSJ0823$+$3231\dotfill  &  348$\pm$56  & 338$\pm$336  & \nodata    & 9.1$\pm$0.3 & 9.2$\pm$0.3 &  13.2$_{-0.5}^{+0.5}$ & 0.39$_{-0.32}^{+0.91}$ \\
 SDSSJ0943$+$3456\dotfill  &  358$\pm$50  & \nodata      & \nodata    & 9.1$\pm$0.2 & 9.6$\pm$0.3 &  13.3$_{-0.5}^{+0.5}$ & 0.43$_{-0.35}^{+1.0}$  \\
 SDSSJ0950$+$0511\dotfill  &  423$\pm$61  & 363$\pm$75   & \nodata    & 9.4$\pm$0.3 & \nodata     &  12.9$_{-0.5}^{+0.7}$ & 0.09$_{-0.07}^{+0.36}$ \\
\enddata
\tablecomments{ \\
Col.(2)--(4): Stellar velocity dispersion in units of km
s$^{-1}$ (corrected for instrumental resolution; \S \ref{subsec:sigma}).
The total uncertainty consists of the measurement error and the systematics dominated by template
mismatch estimated using template stars with different types. \\
Col.(5): Black hole mass in the form of log$(M_{{\rm
BH}}/M_{\odot})$ inferred from $\sigma_{\ast}$, assuming the
$M_{{\rm BH}}$-$\sigma_{\ast}$ relation of \citet{tremaine02}. The
uncertainty on $M_{{\rm BH}}$ listed here is propagated from that
of $\sigma_{\ast}$, which does not include the intrinsic
scatter in the $M_{{\rm BH}}$-$\sigma_{\ast}$ relation \citep[smaller than 0.25--0.3 dex on $M_{{\rm BH}}$;][]{tremaine02}. \\
Col.(6): Virial black hole mass in the form of log$(M/M_{\odot})$.
It is derived based on the width of the broad H$\beta$ line and
the intrinsic quasar continuum luminosity inferred from \loiii\,
(see Table \ref{tab:decomp}) using the formula of
\citet{greene05}. The uncertainty is
dominated by the error of the quasar intrinsic luminosity. \\
Col.(7): Quasar bolometric luminosity in the form of
log$(L/L_{\odot})$ inferred from $L_{{\rm [O\,\,III]}}$. See \S \ref{subsec:eddington} for details.\\
Col.(8): Eddington ratio. The uncertainty is estimated with error
propagation from the uncertainties of $M_{{\rm BH}}$ and $L_{{\rm
Bol}}$, where $L_{{\rm Edd}} \equiv \frac{4\pi G M_{{\rm BH}} m_p
c}{\sigma_{\rm T}}$. }
\end{deluxetable*}

\subsection{Stellar Velocity Dispersion}\label{subsec:sigma}

\begin{figure*}
    \centering
    \includegraphics[width=120mm]{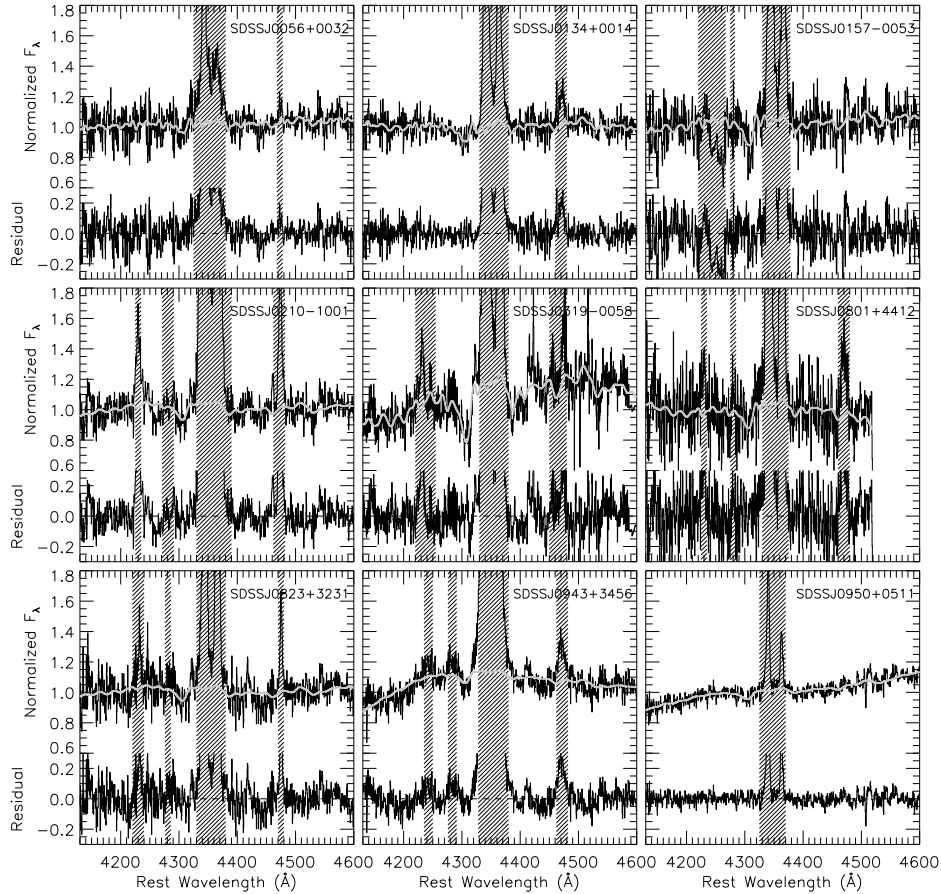}
    \caption{
    Stellar velocity dispersion fits over the G band $\lambda$4304 ${\rm \AA}$ region.
    In each of the nine sub-panels, the normalized spectra (un-smoothed) are presented on the top, along
    with best-fit models over-plotted as gray thick curves. Residuals from the fits
    are displayed on the bottom with null residuals marked as dashed lines.
    The hatched regions are those containing either strong emission lines or
    instrumental chip gaps and have been excluded from the fits.
    The results are listed in Table \ref{tab:sigmafit}.}
    \label{fig:sigmagb}
\end{figure*}

Several stellar absorption features representative of relatively
old populations are discernable in Figure \ref{fig:spec_obscor}.
These include Ca K $\lambda3934$ \angstrom\, and the G band
$\lambda4304$ \angstrom\, seen in most of our targets, and \mgib\
$\lambda$5175 \angstrom\, and Fe $\lambda5270$ \angstrom\,
apparent at least in SDSSJ0157$-$0053. Ca H $\lambda3968$
\angstrom\, is unusable for all objects in our luminous-quasar
sample due to the overlapping strong emission lines of \neiii\
$\lambda3968$ \angstrom\, and H$\epsilon$.

We fit the data with broadened model spectra in pixel space to
include a power-law component to account for possible scattered
quasar light (\S \ref{subsec:scatteredlight}), and mask spectral
regions containing emission lines, detector gaps, and bad pixels.
We use the direct fitting algorithm of \citet{greene06a} and
\cite{ho09}, with a model spectrum constructed by a stellar
template convolved with a Gaussian as an approximation of the
line-of-sight velocity broadening function, plus a power-law
component. We include a power-law component here only to measure
stellar velocity dispersions; the quantification of the
non-stellar continuum is discussed later in \S
\ref{subsec:scatteredlight}. In addition, the summed components
are multiplied by a 3rd-order polynomial in order to model the
difference in continuum shape between data and templates, which
could result from reddening, template mismatch, and/or calibration
uncertainties \citep{greene06a}. Template stars were drawn from
the stellar library of \citet{valdes04}, which have a spectral
coverage of 3460--9464 \angstrom\, with a resolution of $\sim$1
\angstrom\, full width at half maximum (FWHM). This resolution
corresponds to $\sigma \sim$ 15 km s$^{-1}$, much smaller than the
$\sim$ 65 km s$^{-1}$ instrumental resolution of GMOS and SDSS.
The difference between the resolution of the stellar templates and
the instrumental resolution of GMOS has been subtracted in
quadrature from the $\sigma_{\ast}$ measurement. We include
several K stars with different subclasses to account for the
primary contribution from old populations, plus an F2 star and a
G5 star to model potential young stars.

Since the targets are at $z \sim 0.5$, it was impossible to
observe a velocity template star with the identical set-up as the
science targets. We thus adopt the following two-step procedure to
calibrate the instrumental resolution. We first obtained arc-lamp
exposures before and after observing each science target, from
which we measure the instrumental resolution. We then use the
spectroscopic standard star spectrum, taken with Gemini, as a
sanity check by measuring the instrumental resolution both using
our arc-lamp procedure and also from our velocity templates at
much higher resolution. The two measurements for the spectroscopic
standard star agree, lending support to our overall approach of
using arc exposures to calibrate the instrumental resolution.

Spectral fitting was performed using each single star as the
template at first and then with a linear combination of different
types of stars to yield best-fit parameters. There are seven free
parameters in a fit with a single-star template (the center and
width for the Gaussian, the amplitude for the power-law, and four
coefficients for the 3rd-order polynomial). The power-law index is
fixed at $\alpha_{\lambda} = -1.5$; we found that the dispersion
measurements are insensitive to the assumed values of
$\alpha_{\lambda}$ in the range from $-1.5$ to $-1.0$. The
minimization of $\chi^2$ is performed using the nonlinear
Levenberg-Marquardt algorithm implemented by the IDL package
``mpfit''. The uncertainties are dominated by systematics, which
include template mismatch, particularly due to non-solar
abundances in the program galaxies and differing intrinsic widths
for different spectral types \citep[e.g.][]{ho09}. For active
galaxies the situation is further complicated by emission-line
contamination (specifically from Fe and N ions in the \mgib\
triplet and Fe regions; e.g., \citealt{greene06a}). In order to
mitigate these effects, we first measure the local velocity
dispersions over three spectral regions 3700--4020 (Ca K),
4130--4600 (G band), and 5080--5450 (Mg-Fe) \angstrom\, for every
case in which a reasonable fit can be obtained. We define a fit as
``acceptable'' if the reduced $\chi^2 < 8$ and the model traces
stellar absorption features reasonably well as seen by eye. We
found acceptable fits to the G band in all cases, whereas we found
trustworthy fits for six of the nine objects around Ca K and for
only one object over the Mg-Fe region. The Ca K and Mg-Fe regions
suffer more severely from surrounding emission lines than the G
band. Fitting over these two regions is more difficult also due to
the intrinsically smaller equivalent widths (EWs) of \mgib\ and
Fe, and the narrow cores of Ca K resulting from host-galaxy
interstellar absorption. Despite these difficulties, the velocity
dispersions from the Ca K and Mg-Fe regions agree with the G band
results within 1-$\sigma$ uncertainties in all cases. We take the
G band results as fiducial values in the following analysis.

Spectral fittings for the G band over the spectral range
4130--4600 \angstrom\, are displayed in Figure \ref{fig:sigmagb}
for all nine targets, and results from the fits are given in Table
\ref{tab:sigmafit}. The velocity dispersions $\sigma_{\ast}$
listed have been corrected for instrumental resolution (\S
\ref{subsec:obs}). The listed total uncertainty of $\sigma_{\ast}$
contains the measurement error and the estimated systematic
effects, which are mostly dominated by template mismatch. The
measured stellar velocity dispersions range from $\sim$120 to 400
km/s, with a median value of $\sim$290 km/s. The inferred black
hole masses and Eddington ratios are discussed in \S
\ref{subsec:eddington}.

\begin{figure}
    \centering
        \includegraphics[width=65mm]{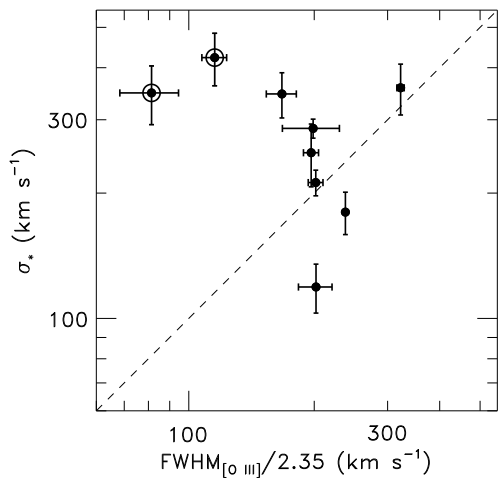}
    \caption{
    Comparison of stellar and gas velocity dispersions. The
    stellar velocity dispersion $\sigma_{\ast}$ is from the G band (Figure \ref{fig:sigmagb}),
    the error of which is dominated by template mismatch. The gas velocity dispersion is
    determined as FWHM/2.35 of \oiii\ 5007, which exhibits asymmetry and/or non Gaussianity for most
    objects in our sample (Figure \ref{fig:spec_obscor}). The uncertainty of FWHM$_{{\rm [O
    \,\text{\tiny III}]}}$ is estimated using the difference between SDSS and GMOS measurements (also see Figure \ref{fig:compare}). The two
    objects marked with open circles have resolved double nuclei in the projected central
    5 arcsec, so the measured $\sigma_{\ast}$ might over-represent that of each component
    (\S \ref{subsec:companion}). There is no correlation between stellar and gas velocity dispersions
    in our sample whether or not the two double-core objects are included \citep[\S \ref{subsec:sigma}; also see][]{greene09}.
    }
    \label{fig:sigmagas}
\end{figure}

The stellar and gas velocity dispersions (estimated as FWHM/2.35
of \oiii) are compared in Figure \ref{fig:sigmagas}. In our sample
there is no correlation between them; the Spearman correlation
coefficient is $\rho = -0.5$ with $P_{{\rm null}} = 0.2$ for the
whole sample and $\rho = -0.2$ with $P_{{\rm null}} = 0.7$ when
excluding the two double-core objects (see below). Other than the
two double-core objects, five out of seven objects have
$\sigma_{\ast} >$ FWHM$_{{\rm [O \,\, \text{\tiny III}]}}/2.35$.
\citet{greene09} also find no correlation in a sample of $\gtrsim
100$ SDSS type 2 quasars with \loiii$/L_{\odot} \sim
10^{8.5}$--$10^{9.1}$. SDSSJ0943$+$3456 (the object at the upper
right corner in Figure \ref{fig:sigmagas}) has the largest gas
velocity dispersion in our sample (FWHM/2.35 of \oiii\ $\sim$ 300
km/s). While its 2-D spectrum shows no strong evidence for
extended gas, it is conceivable that the gas is highly disturbed
since the narrow lines seem to have several components at
different velocities (Figure \ref{fig:spec_obscor}). This might be
further evidence for interaction in addition to its close
companion (see \S \ref{subsec:companion}).

Two caveats must be mentioned here. First, roughly half of
elliptical galaxies in SDSS with the apparently largest velocity
dispersions are in fact close doubles \citep{bernardi06}.
Similarly, our measurements may be biased by additional stellar
components covered by the slit. Indeed, as discussed in \S
\ref{subsec:companion}, SDSSJ0950$+$0511 and SDSSJ0823$+$3231 are
both such double-core systems marginally resolved in our long-slit
observations. As discussed in \S \ref{subsec:eddington}, their
black hole masses (Eddington ratios) listed in Table
\ref{tab:sigmafit} are therefore likely to be over-estimated
(under-estimated). Second, our results have not been corrected for
aperture effects. At redshifts of 0.5, the size of our aperture
(corresponding to $\sim$ 30 kpc) is large enough that the old
stellar populations from which we measure $\sigma_{\ast}$ also
contain those in the disk component, if any, so that the
$\sigma_{\ast}$ (and by extension black hole masses) are likely to
be over-estimated. However, we estimate that this contamination
should be small, considering that most of the host galaxies of
luminous type 2 quasars in our parent sample are likely to be
dominated by ellipticals with de Vaucouleurs light profiles as
seen in the imaging study of a pilot sample \citep{zakamska06}.
For those with disk components, the contamination should still be
moderate since the old stellar populations are predominantly in
stellar bulges.

\subsection{Quantifying Scattered Quasar Light from Broad H$\beta$ in Direct Spectra}\label{subsec:scatteredlight}

\begin{figure}
    \centering
        \includegraphics[width=90mm]{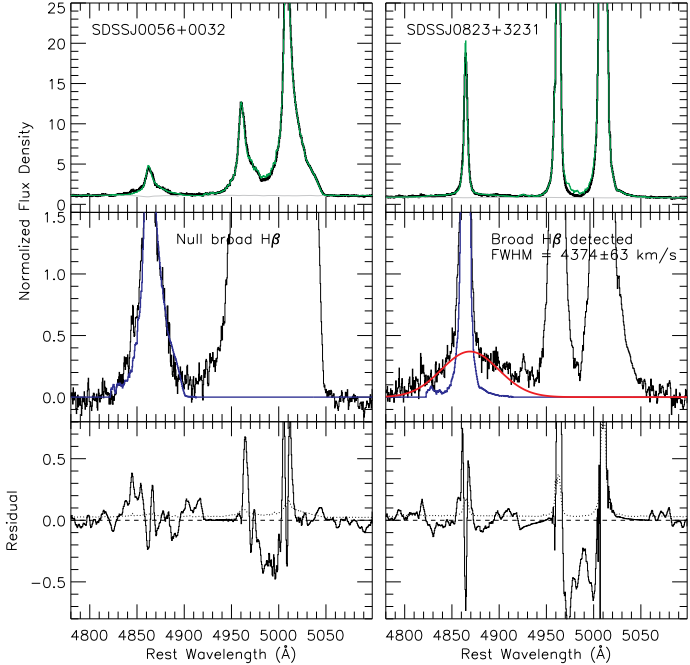}
    \caption{
    Quantifying the broad H$\beta$ component to constrain scattered light contamination of the stellar continuum.
    The two examples represent null
    (SDSSJ0056$+$0032) and significant (SDSSJ0823$+$3231) detections, respectively.
    Each of the narrow emission lines, H$\beta$ and \oiii\ $\lambda\lambda$4959, 5007, is modeled using a template
    combining the blue wing of \oiii\ $\lambda$4959 and the red wing of \oiii\ $\lambda$5007 \citep{reyes08}.
    The broad H$\beta$ component is modeled as a Gaussian.
    {\bf Top row}: spectra are presented over the fitting range 4780$-$5100 \angstrom.
    Data are plotted in black and best-fit models are in green.
    Gray curves show the continuum model which has been subtracted before line fitting.
    {\bf Middle row}: continuum-subtracted spectra zoomed in at the base.
    The model for the broad/narrow H$\beta$ is shown in red/blue.
    {\bf Bottom row}: residuals from the fittings (smoothed with 6-pixel boxcar) are shown ({\it solid}),
    along with 1-$\sigma$ error spectra ({\it dotted}) and null residuals ({\it dashed}).
    }
    \label{fig:broadhbeta}
\end{figure}

Using spectropolarimetry, \citet{zakamska05} detected high
polarization levels in a sample of SDSS type 2 quasars as luminous
as our targets. Significant scattered light has also been revealed
by HST imaging in a somewhat less luminous SDSS type 2 quasar
sample \citep{zakamska06}. Early studies attributed the blue light
frequently observed in the host galaxies of obscured AGN to
scattered light predicted by AGN unification models
\citep[e.g.,][]{antonucci93}. However, it was later realized that
in many cases scattered light explained only part of the blue
continuum \citep[e.g.,][]{kay94,cid95}, as the polarization
fraction observed in the continuum is smaller than that in the
permitted emission lines, requiring an additional source of
non-polarized blue light
\citep[e.g.,][]{goodrich89,tran95,heckman95}. Since there is
strong evidence for the presence of both blue starlight from young
stellar populations and scattered quasar light in the continuum,
we need to determine and remove the contamination from scattered
light before performing stellar population modeling to the
continuum.

To quantify the level of scattered quasar light, we model the
H$\beta$ line as the sum of a narrow and an underlying broad
component from scattered light. Then we make use of the tight
correlation between the H$\beta$ and the continuum luminosities
observed in unobscured AGN \citep[e.g.,][]{greene05} to infer the
corresponding non-stellar continuum. The rms scatter of this
correlation is $\sim 0.2$ dex \citep{greene05} and is added in
quadrature with the measurement error into the uncertainty of our
estimate of the scattered light. The assumption of our approach is
that if broad lines are present with a certain flux, the quasar
continuum should also be present at a level proportional to the
broad emission line flux. H$\beta$ is adopted because it is the
strongest permitted line in the wavelength range covered by the
GMOS spectra.

The procedure to quantify the broad H$\beta$ component is the
following. First a continuum model is constructed by performing a
$\chi^2$ fit to the emission-line-masked spectrum over the range
of 3600--5600 \angstrom\, using multiple instantaneous starburst
templates of \citet{bc03} broadened with the measured stellar
velocity dispersion and a power-law component assuming
$\alpha_{\lambda} = -1.5$. Next we make a four-component fit
(broad H$\beta$, narrow H$\beta$, \oiii\ $\lambda$4959, and \oiii\
$\lambda$5007) to the continuum-subtracted GMOS spectrum over the
range of 4780--5100 \angstrom. We fit each of the three narrow
emission lines non-parametrically with an \oiii -profile model
constructed using the blue wing of \oiii\ 4959 and the red wing of
\oiii\ 5007 \citep{reyes08}. For objects like SDSSJ0943$+$3456,
which appears to have several components with different velocities
in its narrow lines, we do not need to use multiple components to
fit the narrow-line profile, since they are accounted for all at
once using the model constructed from the \oiii\ lines. Unlike
parametric multi-component fits, this non-parametric approach is
free from degeneracies between multiple narrow and broad
components, introducing no additional uncertainty to the
measurement of broad H$\beta$. The assumption here is that the
narrow H$\beta$ component has the same profile as that of \oiii .
Their redshifts are constrained to be the same as they all
originate from the narrow-line region. The broad H$\beta$ is fit
as a single Gaussian. There are seven free parameters in one
fitting (a redshift for the three narrow lines, an amplitude for
each narrow line, and three parameters for the Gaussian). Although
in principle one should iterate the continuum and emission-line
fits, in practice we have found the broad H$\beta$ component to be
insensitive to the details of the continuum fit so that one
iteration is enough.

As pointed out by \citet{reyes08}, it is important to use the
\oiii\ line profiles instead of Gaussians or Lorentzians to fit
the strong narrow emission lines. Almost all the objects in our
luminous-quasar sample have narrow line profiles which deviate
significantly from Gaussian and Lorentzian profiles, often with
strong asymmetries and sometimes double peaks. The specific
profiles of the strong narrow emission lines, if not properly
accounted for, can cause false broad-H$\beta$ detections.

The broad-H$\beta$ measurements are summarized in Table
\ref{tab:decomp}. We unambiguously detect broad H$\beta$
components in four objects; we list 1-$\sigma$ upper limits on
broad H$\beta$ for the other five objects. The uncertainty on the
broad H$\beta$ luminosity is estimated using the 1-$\sigma$
flux-density error spectrum integrated over 4794--4932 \angstrom,
which is equivalent to assuming a Gaussian with a FWHM of 8000
km/s centered at 4863 \angstrom. This is a rather conservative
estimate, considering that the broad-H$\beta$ components detected
in our sample all have FWHM $\gtrsim$ 4000 km/s. Figure
\ref{fig:broadhbeta} presents examples of significant and null
broad-H$\beta$ detections, respectively. The unambiguous
broad-H$\beta$ detections or robust upper limits rely on the high
S/N achieved by our GMOS observations (a median S/N of 56 per
rest-frame 1.0 \angstrom\, over the spectral range of 4841--4881
\angstrom).

\begin{deluxetable}{ccccc}
\tabletypesize{\scriptsize} \tablewidth{0pc}
\tablecaption{Scattered Light Quantification and Continuum
Decomposition Based on Broad-H$\beta$
Measurement\label{tab:decomp}} \tablehead{ \colhead{} & \colhead{}
& \colhead{$L_{5100}^{{\rm scattered}}$} &
\colhead{$\frac{L_{5100}^{{\rm scattered}}}{L_{5100}^{{\rm
obs}}}$} &
\colhead{$\frac{L_{5100}^{{\rm scattered}}}{L_{5100}^{{\rm QSO}}}$} \\
\colhead{Target Name} & \colhead{$L_{{\rm H}\beta}^{{\rm broad}}$}
& \colhead{($10^{43}$ erg/s)} & \colhead{(\%)} &
\colhead{(\%)} \\
\colhead{(1)} & \colhead{(2)} & \colhead{(3)} & \colhead{(4)} &
\colhead{(5)} } \startdata
w/o broad H$\beta$ & & & & \\
\tableline
  & & & &  \\
 SDSSJ0056\dotfill & $<$7.29 & $<$0.74 & $<$ 9 & $<$0.2     \\
 SDSSJ0134\dotfill & $<$7.66 & $<$1.6  & $<$17 & $<$0.2     \\
 SDSSJ0157\dotfill & $<$7.04 & $<$0.45 & $<$ 8 & $<$0.2     \\
 SDSSJ0319\dotfill & $<$7.66 & $<$1.6  & $<$17 & $<$0.2     \\
 SDSSJ0950\dotfill & $<$7.69 & $<$1.7  & $<$ 5 & $<$0.4     \\
  & & & &  \\
\tableline
with broad H$\beta$ & & & & \\
\tableline
  & & & &  \\
 SDSSJ0210\dotfill & 7.84$_{-0.43}^{+0.21}$ & 2.3$_{-1.3}^{+1.2}$ & 37$\pm$22 & 0.2$_{-0.1}^{+0.4}$  \\
 SDSSJ0801\dotfill & 7.70$_{-0.79}^{+0.26}$ & 1.7$_{-1.4}^{+1.2}$ & 45$\pm$36 & 0.3$_{-0.3}^{+0.9}$  \\
 SDSSJ0823\dotfill & 8.16$_{-0.11}^{+0.08}$ & 4.3$_{-0.8}^{+0.8}$ & 55$\pm$12 & 0.4$_{-0.3}^{+0.8}$  \\
 SDSSJ0943\dotfill & 8.09$_{-0.15}^{+0.11}$ & 3.7$_{-1.0}^{+0.9}$ & 28$\pm$ 8 & 0.3$_{-0.2}^{+0.6}$  \\
\enddata
\tablecomments{ \\
Col.(1): Abbreviated target name. \\
Col.(2): Luminosity of the broad-H$\beta$ component in the form of
log$(L/L_{\odot})$ determined from emission-line fits over the
continuum-subtracted GMOS spectra. For objects with null
broad-H$\beta$ detection, 1-$\sigma$ upper limits are given.
The fitting method and error estimation are described in \S \ref{subsec:scatteredlight}. \\
Col.(3): Quasar (scattered light) monochromatic luminosity
$L_{5100}=\lambda L_{\lambda}$ at rest-frame $\lambda=5100$ ${\rm
\AA}$, determined from the broad-H$\beta$ luminosity, using the
$L_{5100}$--$L_{{\rm H}\beta}$ calibration of \citet{greene05}.
The uncertainty contains both that propagated from the
broad-H$\beta$ luminosity and the scatter in the observed
$L_{5100}$--$L_{{\rm
H}\beta}$ correlation. \\
Col.(4): Percentage of scattered quasar light relative to the
total observed luminosity at rest-frame $\lambda=5100$ ${\rm
\AA}$. The uncertainty is propagated from the errors of
$L_{5100}^{{\rm scattered}}$ and $L_{5100}^{{\rm obs}}$. Since
this ratio is a relative quantity, it is independent of the
calibration uncertainty of the GMOS spectra. \\
Col.(5): Percentage of the nuclear quasar light scattered into our
line of sight. The intrinsic $L_{5100}^{{\rm QSO}}$ (unobscured)
is inferred from \loiii\  using the \loiii -$M_{2500}$ calibration \citep{reyes08} and assuming
a spectral index of $\alpha_{\nu} = -0.44$ \citep{vandenberk01}.
The uncertainty contains both that propagated from $L_{5100}^{{\rm
scattered}}$ and the scatter in the \loiii -$M_{2500}$ relation \citep{reyes08}. }
\end{deluxetable}

Using the correlation between H$\beta$ luminosity and
monochromatic continuum luminosity $L_{5100}$ determined from
unobscured AGN by \citet{greene05}, we quantify the amplitude of
the scattered-light component $L_{5100}^{{\rm scattered}}$ (or its
upper limit when un-detected) as listed in Table \ref{tab:decomp}.
We adopt a scattered light spectrum with a spectral index of
$\alpha_{\nu} = -0.44$ \citep{vandenberk01}.

One caveat that must be mentioned here is that we are measuring
the scattered light inside our slit, the contribution of which
could vary with slit position because the scattering can be very
asymmetric \citep[e.g.,][]{zakamska06}. In particular, if the slit
is oriented perpendicular to the axis of the scattering cone, the
scattered light covered by long-slit spectroscopy could be much
smaller than that detected by broad-band polarimetry. In our
sample, only SDSSJ0056$+$0032 has broad-band (observed 480--600
nm) polarization measurement \citep[P. Smith, private
communication;][]{zakamska06}, which indicates a high level of
polarization ($10.2 \pm 1.6$\%) suggesting a significant scattered
light contribution ($> 10$\%) around rest-frame $\sim 3600$
\angstrom. A scattered light fraction of $10$\% at $\sim 3600$
\angstrom\, will translate into $5$\% at $5100$ \angstrom\, (both
in rest-frame) given the spectrum of SDSSJ0056$+$0032, assuming
$\alpha_\nu = -0.44$, which is not in conflict with our 1-$\sigma$
upper limit ($L_{5100}^{{\rm scattered}}/L_{5100}^{{\rm obs}} <
9$\%). However, the estimate of scattered light fraction at $5100$
\angstrom\, based on polarization measurement can be larger than
$5$\% assuming a more typical intrinsic continuum polarization
(e.g., 20\%), and thereby differ from our estimate based on the
strength of broad H$\beta$. This possible discrepancy may be due
to the difference in coverage of the long-slit and broad-band
observations. Indeed, the GMOS slit was oriented within 15\% of
the polarization position angle and was therefore almost
perpendicular to the scattering direction.

\subsection{Stellar Populations}\label{subsec:pop}

Our Gemini spectra cover a spectral range containing various
signposts for both young and old populations. These include: (1)
the 4000-\angstrom\, break and stellar absorption lines (such as
Ca II K\&H at 3934, 3968 \angstrom, the G band at 4304 \angstrom,
the \mgib\ triplet at 5175 \angstrom, and Fe 5270 \angstrom)
indicative of relatively old populations (i.e., with ages $>$ a
few Gyr); (2) the Balmer continuum limit at 3646 \angstrom\, and
the Balmer absorption-line series representing post-starburst
populations (with ages $<$ 1 Gyr); and (3) the broad Wolf-Rayet
emission complexes around $\sim$ 4660 \angstrom\, which trace very
recent starburst activity in the last 5 Myr.

As foreshadowed by \citet{heckman97} in their study of Mrk 477,
one of the most luminous local type 2 quasars, detecting
post-starburst signatures through Balmer absorption becomes more
difficult at high luminosity, because of the strong Balmer
emission lines excited by the AGN. Indeed as seen in Figure
\ref{fig:spec_obscor}, the Balmer absorption series is barely
discernable for most of the nine luminous quasars. The only
exception is SDSSJ0056$+$0032 (which lies at the low-luminosity
end of our sample, and has prominent very young stellar
populations, see below), for which the high-order Balmer
absorption lines (H8, H9, and H10) are visible.

The broad emission complex around \heii\ $\lambda$4686 \angstrom\,
due to Wolf-Rayet (WR) stars, on the other hand, is more prominent
in luminous type 2 quasars. WR stars are very massive evolved
stars that appear only $\sim$2--5 Myr after a burst of star
formation \citep[e.g.][]{vacca92,schaerer98}, whose several km
s$^{-1}$ winds produce the broad emission features. Because of
their short duration, WR stars are an excellent clock for very
recent starburst activity. In addition, the subtypes of WR stars
(e.g. WN or WC) can be inferred from the relative strengths of
different WR features, and the WR luminosity can then be further
translated into the number of WR stars present
\citep[e.g.,][]{smith91,schaerer98}.

\subsubsection{Wolf-Rayet Populations}\label{subsubsec:wr}

\begin{deluxetable}{cccc}
\tabletypesize{\scriptsize}
\tablewidth{0pc}
\tablecaption {
Measurements of Wolf-Rayet Features around \heii
4686\label{tab:wr} }
\tablehead{
\colhead{~~~~~~Target Name\dotfill(1)} &
\colhead{SDSSJ0056} &
\colhead{SDSSJ0134} &
\colhead{SDSSJ0950}
}
\startdata
 $L_{{\rm N\,\text{\tiny V}\,4610}}$~\dotfill(2)                   & \nodata                & \nodata                & $7.06^{+0.07}_{-0.08}$  \\
 ${\rm EW_{N\,\text{\tiny V}\,4610}}$~\dotfill(3)       & \nodata                & \nodata                & $0.71\pm0.12$           \\
 ${\rm FWHM_{N\,\text{\tiny V}\,4610}}$~\dotfill(4)          & \nodata                & \nodata                & $1520\pm280$            \\
 $L_{{\rm N\,\text{\tiny III}\,4640}}$~\dotfill(5)                 & $7.04^{+0.04}_{-0.05}$ & $7.33^{+0.07}_{-0.08}$ & \nodata                 \\
 ${\rm EW_{N\,\text{\tiny III}\,4640}}$~\dotfill(6)     & $2.4\pm0.2$            & $3.9\pm0.7$            & \nodata                 \\
 ${\rm FWHM_{N\,\text{\tiny III}\,4640}}$~\dotfill(7)        & $1920\pm200$           & $2680\pm540$           & \nodata                 \\
 $L_{{\rm He\,\text{\tiny II}\,4686}}$~\dotfill(8)                 & $7.18^{+0.05}_{-0.06}$ & $7.40^{+0.04}_{-0.05}$ & $8.28^{+0.01}_{-0.01}$  \\
 ${\rm EW_{He\,\text{\tiny II}\,4686}}$~\dotfill(9)     & $3.3\pm0.4$            & $4.5\pm0.5$            & $11.9\pm0.3$            \\
 ${\rm FWHM_{He\,\text{\tiny II}\,4686}}$~\dotfill(10)       & $1830\pm250$           & $1600\pm230$           & $6410\pm250$            \\
 $L_{{\rm WR}}$~\dotfill(11)                          & $7.42\pm0.05$          & $7.67\pm0.06$          & $8.31\pm0.01$           \\
 Type~\dotfill(12)                                    & WN6--8                 & WN8                    & WN3--4                  \\
 $N_{{\rm WR}}$~\dotfill(13)                          & $10^5$                 & $10^5$                 & $10^6$                  \\
 $\frac{L_{{\rm WR}}}{L_{{\rm H}\beta}}$~\dotfill(14) & $0.10\pm0.01$          & $0.14\pm0.02$          & $0.80\pm0.03$           \\
\enddata
\tablecomments{ \\
Row(1): Abbreviated target name. \\
Row(2),(5),(8): Luminosity of the WR feature in the form of log($L/L_{\odot}$). \\
Row(3),(6),(9): Equivalent width in \angstrom\ of the WR feature. \\
Row(4),(7),(10): Full width at half maximum in km/s of the WR feature. \\
Row(11): Total luminosity combining all the WR features detected around \heii\ 4686 in the form of log($L/L_{\odot}$). \\
Row(12): Subclass of WR stars estimated based on the intensity ratio
between different WR features according to the classification of \citet{smith96}. \\
Row(13): Number of WR stars present estimated from the luminosity and the subclass. \\
Row(14): Ratio of the WR and H$\beta$ luminosities. This can be
estimated as the lower limit to $N_{{\rm WR}}/N_{\rm O}$ to
first-order approximation \citep[e.g.][]{smith91}. }
\end{deluxetable}

%
\begin{figure*}
    \centering
        \includegraphics[width=130mm]{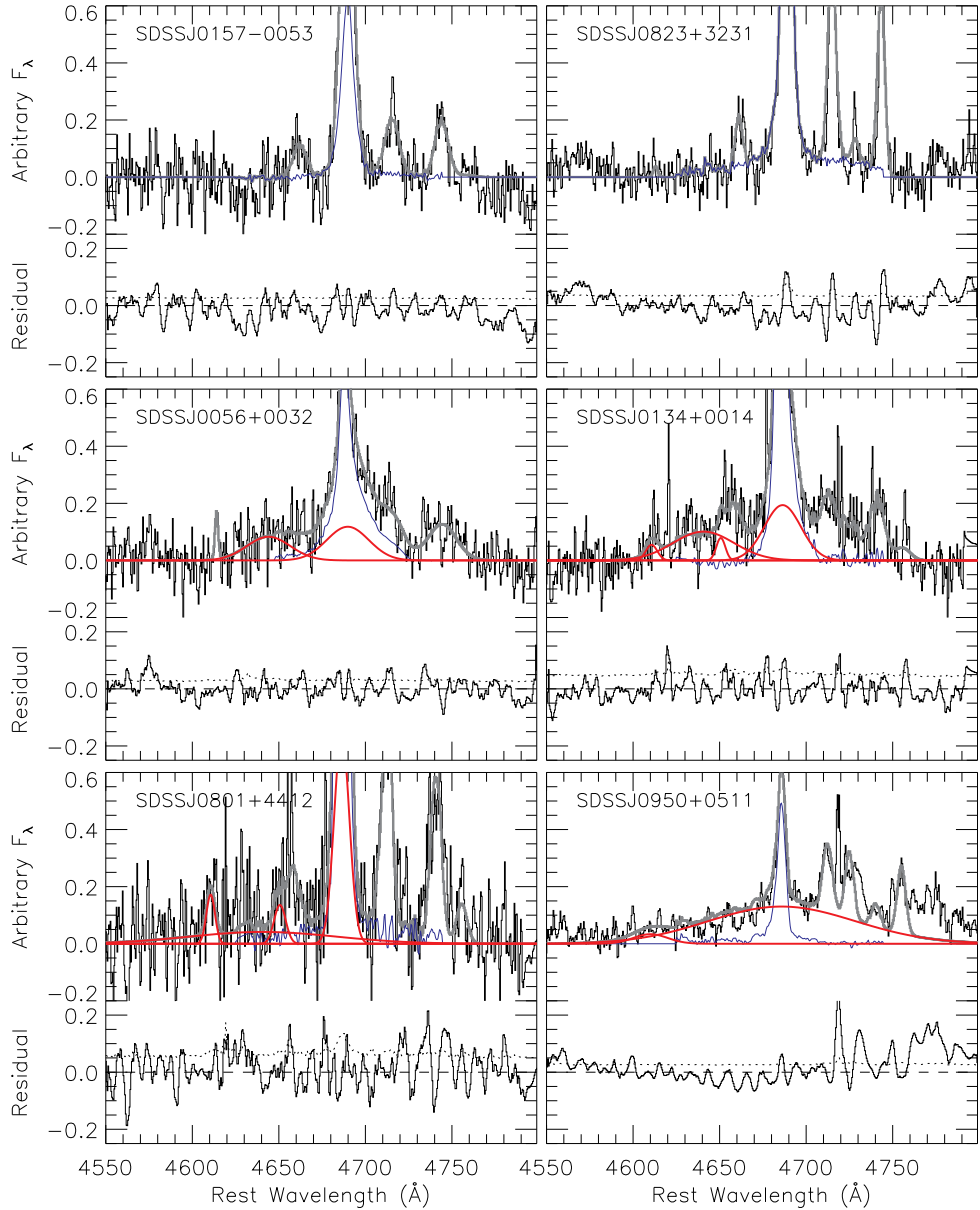}
    \caption{
    Broad Wolf-Rayet (WR) emission complexes around \heii\ $\lambda$4686 \angstrom.
    They include: N {\tiny V} doublet $\lambda$4610 \angstrom, N {\tiny III} $\lambda$4640\angstrom,
    C {\tiny III/IV} $\lambda$4650\angstrom, and \heii\ $\lambda$4686 \angstrom.
    In each example shown, data ({\it black}) and
    best-fit model ({\it dark grey}) are displayed on the top and
    residuals from the fits (smoothed by a 6-pixel boxcar) are shown on the bottom
    along with 1-$\sigma$ error spectra ({\it dotted}) and null residual ({\it dashed}).
    Models for WR features are shown in red.
    Plotted in blue is the model for \heii\ $\lambda$4686, the strongest nebular line
    in this range, which is fit non-parametrically with the observed H$\beta$-line profile.
    Details of our fitting method are presented in \S
    \ref{subsubsec:wr}.
    Two examples of null WR detection are presented in the top row:
    SDSSJ0157$-$0053 has no detectable scattered light, while SDSSJ0823$+$3231 has scattered light detected
    in a broad component to H$\beta$ (Figure \ref{fig:broadhbeta}).
    WR features are detected unambiguously in three targets (SDSSJ0056$+$0032, SDSSJ0134$+$0014, and SDSSJ0950$+$0511).
    The suggestive WR feature is less convincing in SDSSJ0801$+$4412:
    the apparent WR component of \heii\ 4686 could in fact be due to the degeneracy with the nebular component,
    and the detections of the other three WR features are marginal.
    Results from the three clear detections are summarized in Table \ref{tab:wr}.
    }
    \label{fig:WR}
\end{figure*}

In this section, we address the frequency of WR populations in
luminous type 2 quasars. Conspicuous WR signatures in the optical
are less affected by dust obscuration than are UV tracers in
characterizing starburst activity \citep[e.g.][]{kunth99}. WR
populations have been observed in a number of quasar hosts in the
nearby universe, including the famous infrared-luminous galaxy
IRAS 09104 + 4109 \citep{kleinmann88,tran00}, and three luminous
type 2 Seyfert nuclei studied by \citet{gonzalez01} including Mrk
477 \citep{heckman97}. However, the frequency of their occurrence
and their relation to the nuclear activity in luminous quasars
remain important open issues.

In order to quantify WR populations, we fit the broad WR emission
features and the surrounding nebular emission lines simultaneously
to the continuum-subtracted GMOS spectra over 4550--4800
\angstrom. Here we use ``nebular'' to denote the emission lines
that do not arise from WR stellar outflows (which are narrow for
the forbidden lines and could have broad bases due to scattered
light for the permitted lines). Fitting both simultaneously is
required to isolate the WR features \citep{brinchmann08}, but is
only possible with high S/N spectra. The continuum model is
constructed in an identical way as that in the broad H$\beta$ fits
(\S \ref{subsec:scatteredlight}).

The WR features to be detected include: N {\tiny V} $\lambda$4610,
N {\tiny III} $\lambda$4640, C {\tiny III/IV} $\lambda$4650, and
\heii\ $\lambda$4686 \angstrom. They are fit with four Gaussians.
While there are no constraints applied on the widths of these four
WR features, only those broader than the narrow emission lines are
considered as detections. The WR features are broadened as they
arise from stellar outflows and the widths can vary significantly
among individual WR stars.

The surrounding nebular emission lines include [Fe {\tiny III}]
$\lambda$4658, [Fe {\tiny III}] $\lambda$4669, \heii\
$\lambda$4686, [Fe {\tiny III}] $\lambda$4701, [Ar {\tiny IV}]
$\lambda$4711, He {\tiny I} $\lambda$4713, [Ne {\tiny IV}]
$\lambda$4714, [Ne {\tiny IV}] $\lambda$4725, [Ar {\tiny IV}]
$\lambda$4740, and [Fe {\tiny III}] $\lambda$4755 \citep[for the
line list, see e.g.,][]{heckman97,brinchmann08}. The nebular
\heii\ $\lambda$4686 is modeled with the profile of H$\beta$,
which conveniently accounts for the scattered light, if any, to
avoid false WR detection of \heii; each of the other nebular
emission lines is fit with a single Gaussian, the width of which
is fixed to be that from a combined fit of the strong narrow lines
(narrow H$\beta$ and \oiii\ $\lambda\lambda$4959,5007). Gaussian
profiles are adequate for these relatively weak forbidden lines
(from ions of Fe, Ar, and Ne, with Ar lines being the strongest
among them), as we find that given the S/N of our spectra, the
asymmetries in these lines are insignificant. The redshifts of all
the features are fixed to be the same as that of the strong \oiii\
lines.

WR features have been unambiguously detected in three out of the
nine targets, as shown in Figure \ref{fig:WR}. Also displayed in
comparison are two objects with null detections and one object
with only suggestive WR features, which we do not count as a
detection (see figure caption for details). We list results for
the three targets with definitive WR detections in Table
\ref{tab:wr}. Luminosities emitted in each of the WR features and
the total WR luminosity are presented with measurement
uncertainties. The WR luminosities are comparable to or higher
than the most luminous WR galaxies known
\citep[e.g.][]{osterbrock82,armus88}. The inferred subtypes and
the numbers of WR stars are also listed, along with the ratio
between WR and H$\beta$ (narrow) luminosities which can be viewed
to first-order approximation as the ratio between WR and O stars
\citep[e.g.,][]{smith91}. The three targets with unambiguous WR
populations all have WN subtypes \citep{smith96}, similar to Mrk
477 \citep{heckman97}. The C {\tiny III/IV} bump at 4650
\angstrom\, indicative of WC subtypes is not detected in any of
the nine objects. We list the estimates of WN subtypes from the
intensity ratio between different WR features according to
\citet{smith96}. The line widths are consistent with those
expected from the corresponding WN subtypes \citep{smith96}.

\subsubsection{Population Modeling of the Stellar Continuum}\label{subsubsec:stellarpop}

\begin{figure}
    \centering
   \includegraphics[width=70mm]{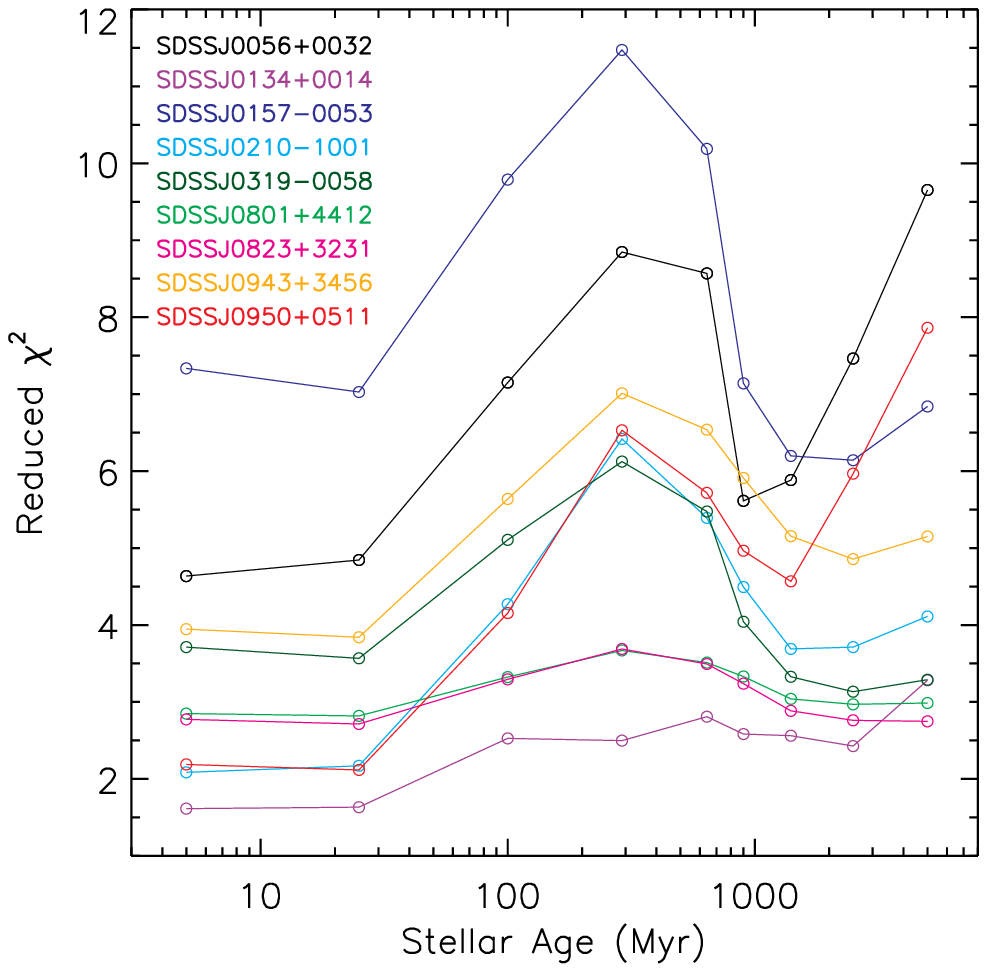}
    \caption{Reduced $\chi^2$ from stellar population fitting using
    a single-age instantaneous starburst model. Contamination from
    scattered quasar light has been excluded before the fitting.
    Results for different targets are color-coded as labeled on the plot.
    See \S \ref{subsubsec:stellarpop} for more discussion.}
    \label{fig:singleage}
\end{figure}
\begin{figure*}
    \centering
        \includegraphics[width=130mm]{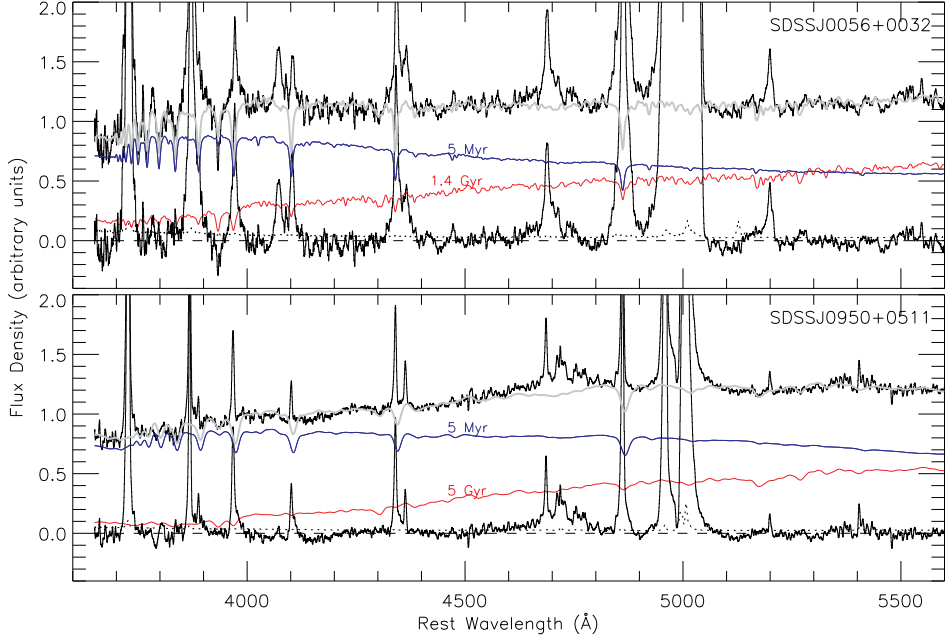}
    \caption{Stellar population analysis results for objects without
    detectable scattered quasar light. Two examples are shown here,
    for objects having the smallest (SDSSJ0056$+$0032) and
    largest (SDSSJ0950$+$0511) $\sigma_{\ast}$ measurements in our
    sample. The normalized spectra (smoothed with a 6-pixel boxcar)
    are shown as thin curves, and best-fit continuum models
    are plotted as thick gray curves. Residuals from the fits are
    displayed on the bottom along with 1-$\sigma$ error spectra ({\it dotted})
    and null residuals ({\it dashed}). The two fit stellar populations are also shown, labelled with stellar age.
    Young populations ($<$ 0.1 Gyr) contribute a considerable fraction of the total stellar luminosity.
    See \S \ref{subsubsec:stellarpop} for more information.
    }
    \label{fig:specdecomp}
\end{figure*}
\begin{figure*}
    \centering
        \includegraphics[width=130mm]{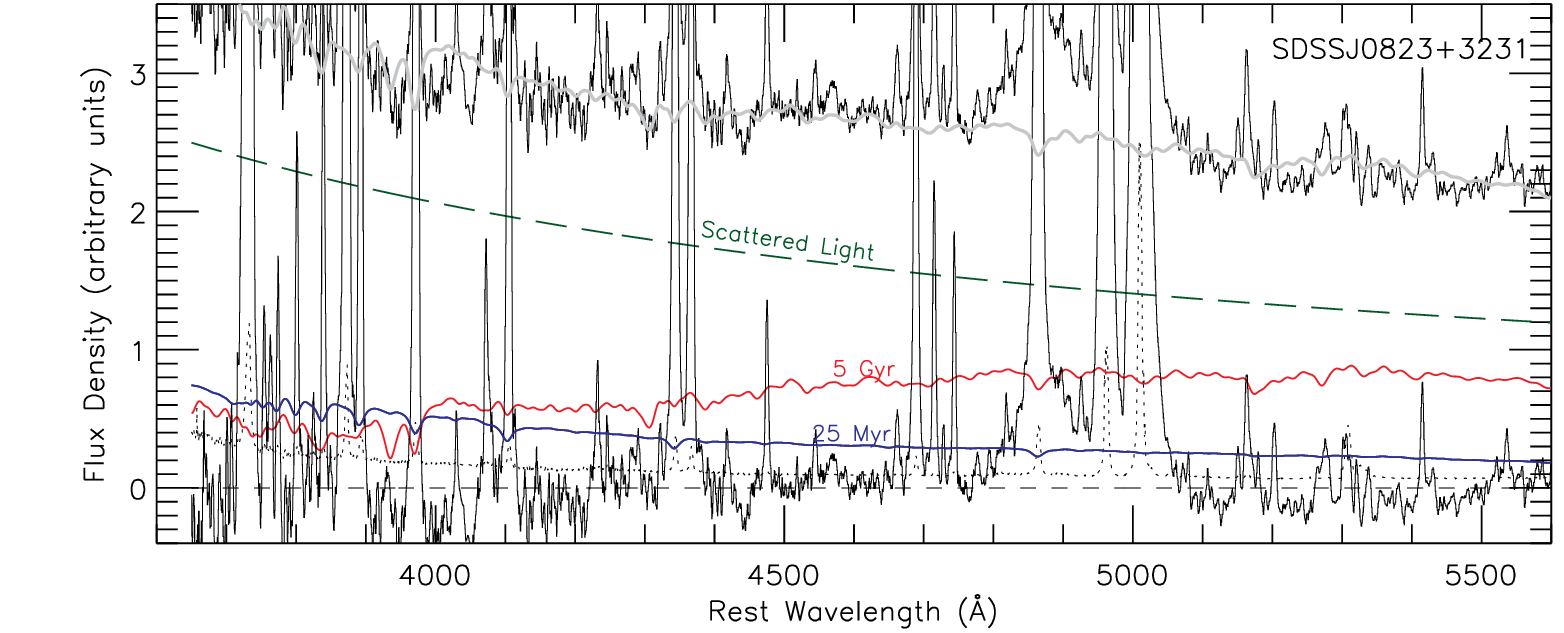}
    \caption{As in Figure \ref{fig:specdecomp}, but for an object
    with detected scattered quasar light as apparent from the detection of a
    broad component to H$\beta$. This object has the highest scattered-light contribution
    to the monochromatic luminosity $L_{5100}$ ($\sim$55\%; Table
    \ref{tab:decomp}) in our sample. The notation is the same as in
    Figure \ref{fig:specdecomp}; the scattered light model is plotted as a long-dashed curve. See \S
    \ref{subsubsec:stellarpop} for more information.
    }
    \label{fig:specdecomp_pl}
\end{figure*}

We now fit stellar population models to the stellar continuum to
get a more complete view of the mix of ages in the host galaxies.
First we fit the stellar continuum using a linear combination of
nine instantaneous starburst models with ages of 0.005, 0.025,
0.10, 0.29, 0.64, 0.90, 1.4, 2.5, and 5.0 Gyr from \citet{bc03},
after masking emission lines and detector chip gaps.   We use
those nine age grids to sample the whole history of a $z \sim 0.5$
galaxy, having in mind a model in which the bulk stellar
population in a galaxy was built up in multiple starburst events
at different epochs.  For objects that have broad H$\beta$
detections (\S \ref{subsec:scatteredlight}), a model of scattered
quasar light is subtracted assuming $\alpha_{\nu} = -0.44$
\citep{vandenberk01}. In producing the instantaneous starburst
model, we assume ``Padova1994'' stellar evolutionary tracks
\citep{bc03} and a Chabrier initial mass function
\citep[IMF;][]{chabrier03} with a low and high mass cutoff of 0.1
and 100 $M_{\odot}$, respectively. We adopt solar metallicity in
the baseline models; assuming super-solar metallicity would result
in smaller age estimates whereas assuming subsolar metallicity
would result in larger ages. Template spectra are broadened by the
measured stellar velocity dispersions (\S \ref{subsec:sigma}).

Fitting was performed over the rest-frame spectral range of
3650--5600 \angstrom. Fitting instead over the range 3600--5600
\angstrom\, results in a slightly higher contribution from young
stellar populations for the objects with the largest \loiii\, in
our sample. This is most likely due to the Balmer continuum
emission blue-ward of 3646 \angstrom\, from the quasar which could
bias stellar ages towards younger values.

\begin{deluxetable}{crccccc}
\tabletypesize{\scriptsize} \tablewidth{0pc} \tablecaption { Mass
and Age Estimates of Stellar Populations \label{tab:mass} }
\tablehead{ \colhead{} & \colhead{} & \colhead{$t_1$} &
\colhead{$f_{L_{5100}}$} & \colhead{$f_{M_{\ast}}$} &
\colhead{$t_2$}  &
\colhead{SFR}   \\
\colhead{Target Name} & \colhead{$M_{\ast}$} & \colhead{(Gyr)} &
\colhead{(\%)} & \colhead{(\%)} & \colhead{(Gyr)} &
\colhead{($M_{\odot}$yr$^{-1}$)} \\
\colhead{(1)} & \colhead{(2)} & \colhead{(3)} & \colhead{(4)} &
\colhead{(5)} & \colhead{(6)} & \colhead{(7)} } \startdata
w/o broad H$\beta$ & & & & & &  \\
\tableline
  & & & & & &  \\
 SDSSJ0056\dotfill & 10.0 & 0.005 & 53 &    2 &   1.4  &  36  \\
 SDSSJ0134\dotfill & 10.3 & 0.005 & 46 &    1 &   2.5  &  42  \\
 SDSSJ0157\dotfill & 10.3 &   1.4 & 44 &   33 &   2.5  &  \nodata  \\
 SDSSJ0319\dotfill & 10.8 & 0.025 & 19 &    1 &   5.0  &  20  \\
 SDSSJ0950\dotfill & 10.9 & 0.005 & 64 &    1 &   5.0  &  150  \\
  & & & & & &  \\
\tableline
with broad H$\beta$ & & & & & &  \\
\tableline
  & & & & & &  \\
 SDSSJ0210\dotfill &  8.3 & 0.005 & 100 & 100 &   5.0  &  44   \\
 SDSSJ0801\dotfill & 10.0 & 0.005 &  38 &   2 &   5.0  &  44   \\
 SDSSJ0823\dotfill & 10.3 & 0.025 &  23 &   1 &   5.0  &   8   \\
 SDSSJ0943\dotfill & 10.2 & 0.025 &  80 &  11 &   5.0  &  68   \\
\enddata
\tablecomments{ \\
Col.(1): Abbreviated target name. \\
Col.(2): Estimate of total stellar mass in the form of log$(M_{\ast}/M_{\odot})$. \\
Col.(3): Age estimate of the younger one of the two stellar
populations which have the
highest contribution to $L_{5100}$. \\
Col.(4): Percentage contribution of the population in Column 3 to $L_{5100}$ after scattered light is subtracted. \\
Col.(5): Percentage contribution of the population in Column 3 to stellar mass. \\
Col.(6): Age estimate of the older one of the two stellar populations which have the highest contribution to $L_{5100}$. \\
Col.(7): Average star formation rate estimate in the past $<0.1$
Gyr. }
\end{deluxetable}

We have also tested fitting the stellar continuum using only one
instantaneous starburst template to the scattered-light subtracted
spectra (using scattered-light models constructed with either
detected broad-H$\beta$ components or upper limits). The reduced
$\chi^2$ as a function of age is shown in Figure
\ref{fig:singleage}. For each object there are two ages at which
the reduced $\chi^2$ reaches a local minimum. In almost all cases,
they represent two populations, one relatively young ($<$0.1 Gyr)
and one relatively old ($>$ 1 Gyr), which dominate the starlight.
Fitting a single-age model prefers a young population ($<$0.1 Gyr)
over an older ($>$1 Gyr) one in most cases. Thus, the data cannot
be explained by pure old populations plus scattered light; there
is a substantial contribution from young populations. The ages of
the young populations that we find are significantly smaller than
the typical post-starburst ages ($\gtrsim$ 1--2 Gyr) estimated in
host galaxies of quasars with lower luminosities \citep[][see \S
\ref{subsec:comparison}]{kauffmann03,canalizo07,bennert08}.

Similarly, in the nine-component fittings, we find that two
components dominate the fits in almost every case, one relatively
old ($> 1$ Gyr) and one relatively young ($< 0.1$ Gyr), so that we
redo the fits with just these two.  The two components with the
highest weights in a nine-component fit also produce the best fit
when we allow the ages to vary (among the adopted nine age grids)
in a two-component fit, which justifies our choice. Two-population
fitting produces a reduced $\chi^2$ at least 10\% smaller than
that when we fit the data with a single population, while adding a
third population does not significantly improve the fits.

Figure \ref{fig:specdecomp} displays two examples of population
synthesis results for objects without detectable scattered quasar
light, and the fit for the object with the highest scattered light
fraction is illustrated in Figure \ref{fig:specdecomp_pl}. In
Table \ref{tab:mass} we list the estimated total stellar masses,
ages, and fractional contributions to stellar luminosity and mass
of the two populations. The uncertainty of the age estimates can
be inferred from the population model grids we use. For example,
if the listed best-fit age is 0.005 Gyr, the next model examined
was 0.025 Gyr, so its uncertainty can be estimated as $< 0.025$
Gyr. We find that eight of our nine targets contain a very young
($< 0.1$ Gyr) and an old ($> 1$ Gyr) population; in half of these,
the age estimates for the young components are $< 25$ Myr. The one
object which does not have a considerable contribution from a $<
0.1$ Gyr population (SDSSJ0157$-$0053) has an age estimate for its
younger population of 1.4 Gyr. The young stellar populations in
the three objects with Wolf-Rayet populations (\S
\ref{subsubsec:wr}) all have best-fit ages of 5 Myr for the young
population, and this population contributes $> 50$\% of $L_{5100}$
in all three cases. This age estimate from population modeling of
the stellar continuum is in good agreement with the detection of
Wolf-Rayet populations, even though they were constrained
independently.

In Table \ref{tab:mass} we also list the average star formation
rates (SFR) in the past $<0.1$ Gyr estimated from the stellar
masses contained in the young stellar population divided by its
estimated age. The inferred median SFR in our sample is $\sim 44$
$M_{\odot}$ yr$^{-1}$ (uncorrected for extinction). This is
smaller than the median SFR ($\sim 87$ $M_{\odot}$ yr$^{-1}$)
inferred from the IR luminosities of the \citet{zakamska08} sample
of 12 SDSS type 2 quasars (with a median \loiii\, of
$10^{9.4}L_{\odot}$, comparable to that of our sample) assuming
the calibration of SFR--$L_{FIR}$ for starburst from
\citet{kennicutt98}. In particular, SDSSJ0056$+0032$ is also in
the \citet{zakamska08} sample and our estimate of its SFR from
population synthesis ($\sim 36$ $M_{\odot}$ yr$^{-1}$) is smaller
than that inferred from its IR luminosity ($\sim 137$ $M_{\odot}$
yr$^{-1}$). The discrepancy in its two SFR estimates suggests an
extinction level of $\sim 1.5$ mag.

Several caveats must be considered here. First, the results are
all based on spectra uncorrected for internal reddening or
extinction so that the inferred contribution from young stellar
populations and the stellar luminosity and mass estimates could be
underestimated. If scattering is by dust, then the SED of the
scattered light may be bluer (or redder) than the intrinsic
$\alpha_{\nu} = -0.44$, leading to an overestimate (or
underestimate) of the young population contribution. The effects
of dust in the narrow-line region are discussed in \S
\ref{subsec:dust}. Third, while we have subtracted scattered
quasar light based on the detection of broad H$\beta$ (\S
\ref{subsec:scatteredlight}), there still might be some
underestimated scattered light which could contaminate the stellar
continuum if the broad component had width $\gg 8000$ km s$^{-1}$.
Nevertheless, the effect of this possibility on our results is
likely to be small, considering broad-line quasars with such
extreme line widths are quite rare. Finally, since our targets
were selected from luminous type 2 quasars, the results on host
galaxies may not necessarily apply to un-reddened luminous type 1
quasars, since they may trace different evolutionary phases
(discussed below).

\begin{figure}
    \centering
        \includegraphics[width=60mm]{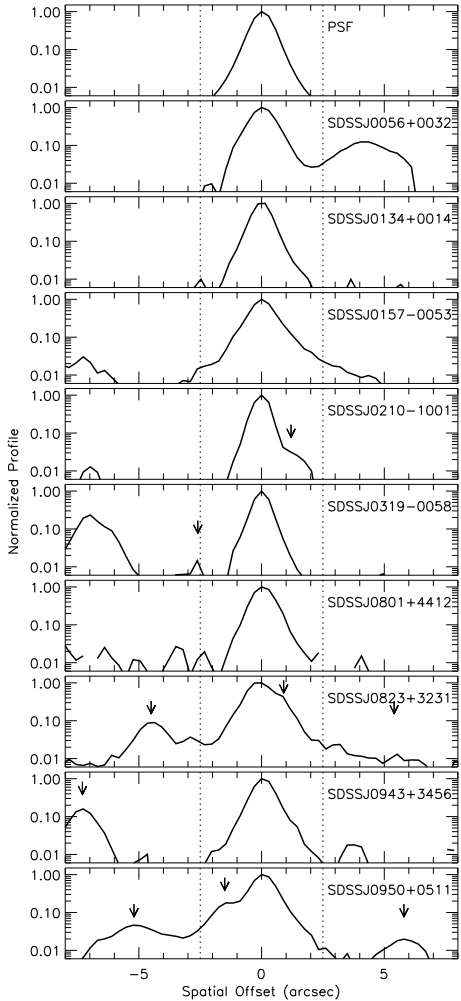}
    \caption{Spatial profiles of all objects in our sample and
    of the standard star (top panel marked ``PSF'') as measured
    from our long-slit observations. These spatial profiles have been
    normalized and centered on the central most luminous peaks,
    combined over the whole observed wavelength range in a
    single exposure for each object.
    The positions of double cores, close companions, and
    extended emission-line regions are indicated by arrows.
    1 arcsec corresponds to 6 kpc at redshift of 0.5 in the assumed cosmology.
    The component at $\sim 4$ arcsec from SDSSJ0056$+$0032 and
    that at $\sim -7$ arcsec from SDSSJ0319$-$0058 have
    line-of-sight velocity offsets $>10^3$ km/s
    relative to the central component; we do not consider them
    to be physically related companions. The spectrum of the neighboring component in
    SDSSJ0210$-$1001 is too noisy
    to allow a redshift to be measured.
    Our long-slit observations show that at least four of the nine targets
    have double cores and close companions or extended emission-line regions.
    The profiles are consistent among different exposures for the same object.
    Figure \ref{fig:2d} displays 2D spectra for the two objects
    with double cores (SDSSJ0823$+$3231 and SDSSJ0950$+$0511).
    See Table \ref{tab:companion} for measurements of the neighboring components.
    }
    \label{fig:spatial}
\end{figure}
\begin{figure}
    \centering
        \includegraphics[width=42mm]{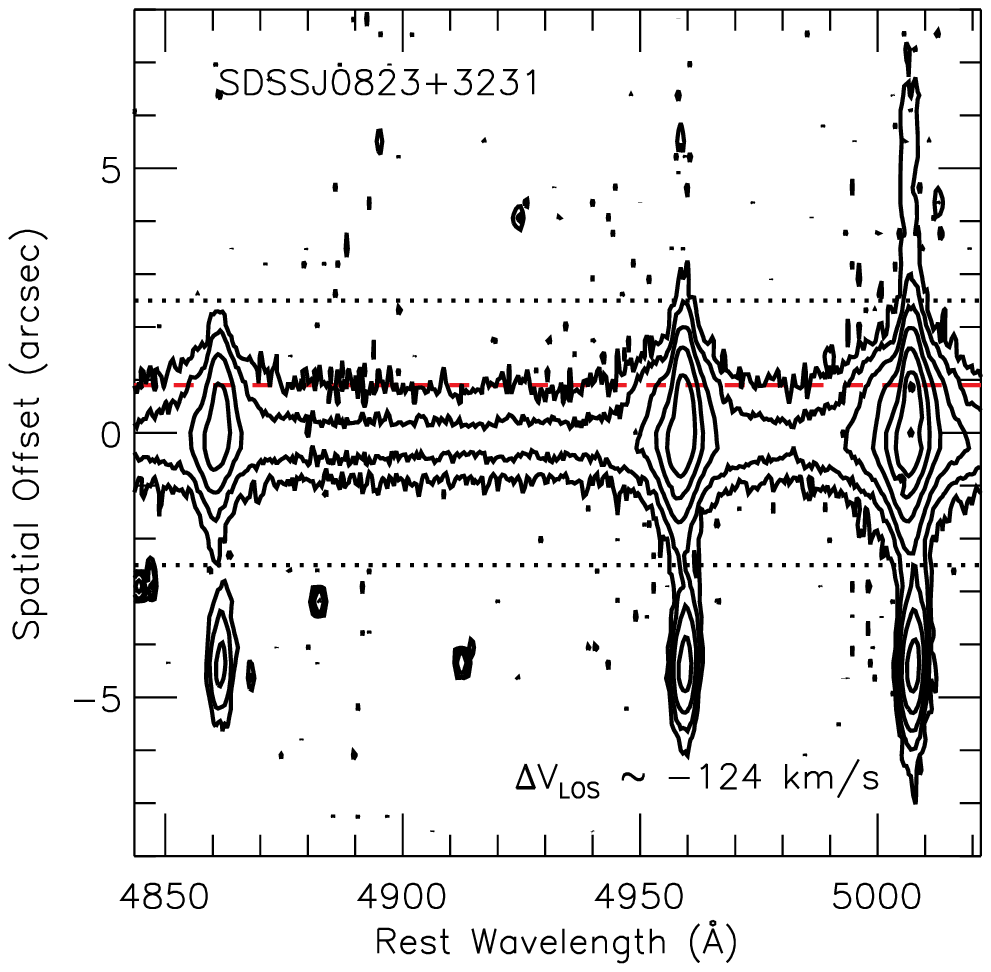}
        \includegraphics[width=42mm]{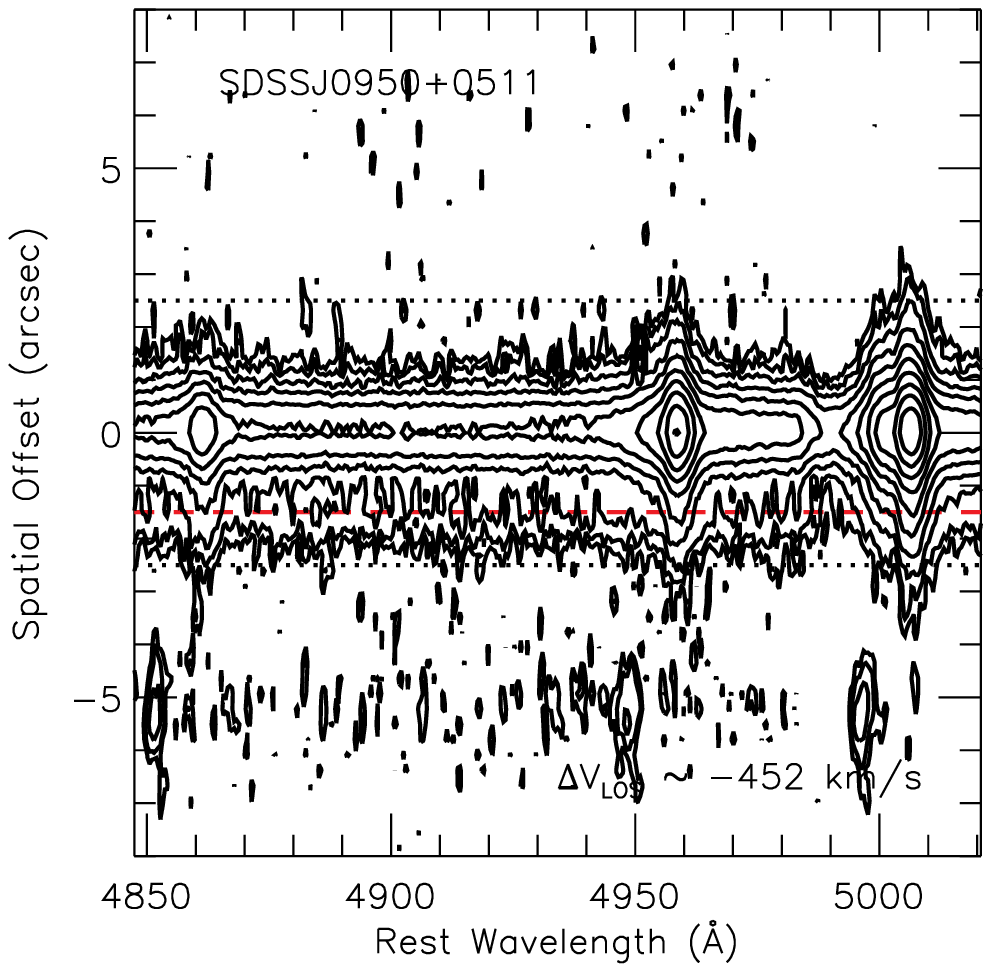}
    \caption{2D spectra of the 2 objects in our sample containing double cores.
    The spectra shown are before flux calibration
    and are uncorrected for telluric absorption, which causes the trough
    at $\sim$ 4990 \angstrom\, in SDSSJ0950$+$0511. Contours
    are plotted in log-linear scales; the lowest and highest
    levels are 10$^{1.50}$ (10$^{1.25}$) and 10$^{4.50}$
    (10$^{3.50}$) counts with a spacing of 0.50 (0.25) dex, amounting to 7 (10) levels in
    total for SDSSJ0823$+$3231 (SDSSJ0950$+$0511). Spectra have been centered on the most luminous
    peaks along the spatial direction.
    The position of the second central component is marked with a red dashed line.
    Dotted lines mark the position of the aperture used to extract 1D spectra.
    In SDSSJ0823$+$3231, the second central
    component is most prominent in emission lines whereas in
    SDSSJ0950$+$0511 it is most prominent in the continuum.
    There are additional companions and extended emission line regions with spatial offsets
    $\Delta S_{{\rm LOS}}$$\sim \pm5$ arcsec. The LOS velocity offsets of these components are indicated on the
    plot. The two neighbor components of SDSSJ0950$+$0511 have clear continua and
    much smaller \oiii /H$\beta$ ratios ($\gtrsim 1$)
    than that in the central nucleus; these ratios are more characteristic of star-forming H II regions.
    In contrast, the two components of SDSSJ0823$+$3231 exhibit almost the same \oiii /H$\beta$,
    \oiii /\oii, and [Ne {\tiny V}]/H$\beta$ ratios as those of the nucleus,
    characteristic of AGN excitation. The component at a projected spatial
    offset of $\sim -5$ arcsec has a clear continuum detection
    (although it is not apparent in the plot due to the scaling) whereas that at $\sim 5$
    arcsec has a weak continuum barely detected redward of $\sim 5100$ \angstrom.
    }
    \label{fig:2d}
\end{figure}

\subsection{Double Cores, Companions, and Extended Emission Line Regions}\label{subsec:companion}

In Figure \ref{fig:spatial} we show spatial profiles of the nine
targets constructed by collapsing the spectra over the whole
observed wavelength range, centered on the most luminous peak
along the spatial direction. The spatial profile of the
photometric standard star is also shown for comparison. While in
many objects the central spatial profiles are consistent with a
single point-spread function (PSF), those of at least two objects
(SDSSJ0823$+$3231 and SDSSJ0950$+$0511) deviate significantly from
a single PSF and exhibit evidence for two components within the
central 5 arcsec (what we call ``double cores''). Arrows mark all
the cases of double cores, companions, and extended emission line
regions (we refer to the latter two cases as ``companions'' in
general) that are physically associated with the central
components. The physical association is verified by determining
the redshifts of companions using either emission or absorption
lines.  The 2D spectra of the two objects with double cores
centered on H$\beta$ and \oiii\ along the wavelength direction are
shown in Figure \ref{fig:2d}.

There are several possibilities concerning the nature of the
companions, which include: (a) a cloud of gas which either is
photoionized by the central nucleus and/or heated by shocks, or
whose dust is reflecting nuclear light; (b) a region of active
star formation; and (c) a companion galaxy which either is merging
with the central nucleus or is a leftover from a past interaction
event, or a small satellite which is not interacting with the
nucleus at all (at the time of observations). Discriminating among
these possibilities is useful to assess the significance of galaxy
interactions in triggering luminous quasar activity. We list the
projected spatial offsets, line-of-sight velocity offsets, SDSS
magnitudes, and several diagnostic emission-line measurements of
the companions and double cores in Table \ref{tab:companion}. A
companion to SDSSJ0056$+$0032 and a companion to SDSSJ0319$-$0058
with relative radial velocities $> 1000$ km/s are not listed.

\begin{deluxetable*}{ccccccccccccccccc}
\tabletypesize{\scriptsize} \tablewidth{0pc} \tablecaption{
Companions, Double Cores and Extended Emission-line
Regions\label{tab:companion} } \tablehead{ \colhead{~~~~~~Target
Name\dotfill(1)} & \colhead{} & \multicolumn{3}{c}{SDSSJ0319} &
\colhead{} & \multicolumn{3}{c}{SDSSJ0823} & \colhead{} &
\multicolumn{3}{c}{SDSSJ0943} & \colhead{} &
\multicolumn{3}{c}{SDSSJ0950}
\\
\cline{3-5} \cline{7-9} \cline{11-13}  \cline{15-17} } \startdata
 $\Delta_{S}$~(kpc)\dotfill(2)                    &  & & $-18$ & &   & $5$     & $-25$   & $30$    &  & &    $46$ & &  & $-9$    &   $-32$  &  $37$
\\
 $\Delta_{v}$~(km/s)\dotfill(3)                   &  & &$-198$ & &   & \nodata & $-124$  & $-170$  &  & &   $-96$ & &  & $-383$  &  $-452$  &  $87$
\\
 $r$\dotfill(4)                                   &  & &\nodata& &   & \nodata & $22.75$ & \nodata &  & & $21.30$ & &  &\nodata  & $22.14$  & \nodata
\\
 EW$_{{\rm [O\,\, \text{\tiny III}]}}$~(${\rm \AA}$)\dotfill(5) &  & & $4.5$ & &   & \nodata &   $682$ & $52.5$  &  & & \nodata & &  &\nodata  &  $15.9$  & $11.2$
\\
 FWHM$_{{\rm [O\,\, \text{\tiny III}]}}$~(km/s)\dotfill(6)      &  & & $167$ & &   & \nodata &   $212$ &  $243$  &  & & \nodata & &  &\nodata  &   $332$  &  $640$
\\
 log($L_{{\rm [O\,\, \text{\tiny III}]}}/L_{\odot})$~\dotfill(7)&  & &$7.45$ & &   & \nodata & $9.08$  & $7.88$  &  & & \nodata & &  &\nodata  &  $7.93$  & $7.43$
\\
 ${\rm [O\,\, \text{\tiny II}]/[O\,\, \text{\tiny III}]}$~\dotfill(8)         &  & &\nodata& &   & \nodata & $0.10$  & $0.43$  &  & & \nodata & &  &\nodata  &   $1.9$  &  $1.7$
\\
 ${\rm H\beta/[O\,\, \text{\tiny III}]}$~\dotfill(9)            &  & & $1.8$ & &   & \nodata & $0.12$  & $0.09$  &  & & \nodata & &  &\nodata  &  $0.48$  & $0.15$
\\
\enddata
\tablecomments{ \\
Row(1): Abbreviated target name. \\
Row(2): Projected spatial offset (physical distance) with respect
to the central nucleus.
The plus or minus sign follows the definition of Figure \ref{fig:spatial}. \\
Row(3): Line-of-sight velocity shift relative to the central nucleus. \\
Row(4): $r$-band model magnitude from SDSS when available. Several
components
do not have SDSS photometry because they are too faint. \\
Row(5): Rest-frame equivalent widths of \oiii\ $\lambda$5007
measured for the components containing line emission. The spectrum
of the component in SDSSJ0943 is free of emission lines.
The double cores of SDSSJ0823 and SDSSJ0950 are not well spatially resolved (namely within the central 5 arcsec). \\
Row(6): Full widths at half maximum of \oiii\ $\lambda$5007 corrected for instrumental resolution. \\
Row(7): Line fluxes of \oiii\ $\lambda$5007. \\
Row(8): Emission-line ratio $\frac{F_{{\rm [O\,
 \text{\tiny II}]}\lambda\lambda3727,3729}}{F_{{\rm [O\,  \text{\tiny III}]}\lambda5007}}$
characterizing the ionization
parameter. There is no measurement for the component in SDSSJ0319 since its \oii\ $\lambda$3727 line overlaps the GMOS chip gap. \\
Row(9): Emission-line ratio $\frac{F_{{\rm H\beta}}}{F_{{\rm [O\,
 \text{\tiny III}]}\lambda5007}}$. }
\end{deluxetable*}

The companions in SDSSJ0319$-$0058 and in SDSSJ0950$+$0511 are
likely to be regions of active star formation, given their
diagnostic emission line ratios \oii /\oiii\ and/or
H$\beta$/\oiii\ \citep[e.g.,][]{osterbrock89}. For the companion
in SDSSJ0319$-$0058, there is no evidence for disturbance or
interaction with the central nucleus in its 2-D spectrum (e.g.,
multiple components at different velocities or tidal tails). In
the double-core system SDSSJ0950$+$0511, it is likely (see below)
that galaxy interaction is responsible for both the quasar and the
young starburst activity, and that the luminous type 2 quasar is
at an early stage of interaction. The companion at $\sim -32$ kpc
has a line-of-sight (LOS) velocity offset (relative to the more
luminous core) similar to that of the less luminous core (Figure
\ref{fig:2d}, right panel). This suggests that the companion is
physically associated with the less luminous core and might have
been stripped off as a result of the merger in the nucleus (which
is marginally resolved in this case).

The companion in SDSSJ0943$+$3456 is an absorption-line galaxy. It
is likely to be a relic core from a past interaction event with
the central galaxy, because it has a best-fit stellar age of 5 Gyr
from stellar population synthesis, which is the same as the older
population in the circum-nucleus stellar component (Table
\ref{tab:mass}). We have detected several stellar absorption
features (including Ca K\&H, the G band, \mgib, and Fe 5270) in
its spectrum and measured its $\sigma_{\ast}$ to be $295\pm9$
km/s.

According to the diagnostic line ratios \oii /\oiii\ and
H$\beta$/\oiii\, the companions in SDSSJ0823$+$3231 are likely to
be gas clouds either photoionized by the central nucleus and/or
heated by shocks, or their dust is reflecting quasar light, rather
than regions of active star formation. They are unlikely to be
galaxies, as there are no stellar features detected in the
continua. The non-stellar continua are most likely scattered
quasar light. The emission lines seem to be from gas photoionized
by the central nucleus. The absence of broad lines, and the large
EWs of the narrow emission lines suggest that these lines are not
due to scattering from gas clouds within the ionizing cones;
scattered light from gas clouds outside the ionizing cones are
unlikely to produce the bulk of the observed luminosities of the
companions. Shock heating also seems unlikely as the FWHMs of the
emission lines are small. Since SDSSJ0823$+$3231 is a double-core
system, these gas clouds might be gas shreds resulting from the
on-going merger.

In summary, our long-slit spectroscopy reveals that at least four
of our nine targets contain double cores and/or physically
associated companions. Three of the four objects show evidence for
galaxy interactions which may be responsible for the quasar and
starburst activity. The companions in these systems have various
origins, including a leftover galaxy core dominated by old stellar
populations, merging galaxies with active star formation, and
perhaps gas clouds shredded by the merger. While we oriented the
slit to cover potential companions seen in the imaging data, our
spectra do not resolve scales $\lesssim$ 1 arcsec so that the
interaction fraction we find is a lower limit to the true value.

\section{Implications and Discussion}\label{sec:discuss}

In this section we present implications and discussion on our
results. We infer black hole masses and Eddington ratios in \S
\ref{subsec:eddington}, and discuss the origins of the blue
continua in luminous type 2 quasars in \S \ref{subsec:fc2}, with a
prescription for the scattered quasar light in \S
\ref{subsubsec:prescription} and the starburst-quasar link in \S
\ref{subsec:starburst}. Extinction and its effects on our results
are provided in \S \ref{subsec:dust}. In \S
\ref{subsec:comparison}, we compare our results with those from
other quasar host-galaxy studies, both obscured (\S
\ref{subsubsec:type2}) and unobscured (\S \ref{subsubsec:type1}).

\subsection{Black Hole Mass and Eddington Ratio}\label{subsec:eddington}

We infer black hole masses from the measured stellar velocity
dispersions (\S \ref{subsec:sigma}) based on the correlation
between bulge stellar velocity dispersion and dynamical black hole
masses observed in local inactive galaxies
\citep{ferrarese00,gebhardt00}, using the calibration of
\citet{tremaine02}. The calibration for local inactive galaxies
does not necessarily directly apply to active galaxies
\citep[e.g.][]{greene06b}, considering that black holes in active
galaxies are still in growth \citep[e.g.,][]{ho08b}. Furthermore,
while our targets only have moderate redshifts $z \sim 0.5$, there
could be non-negligible redshift evolution in the $M_{{\rm
BH}}$-$\sigma_{\ast}$ relation \citep[e.g.][]{woo08,canalizo08}.
For objects with detected broad H$\beta$, we examine this redshift
evolution in \S \ref{subsubsec:virmass}. We estimate accretion
rates bearing in mind these uncertainties. Adopting \loiii\, as a
proxy for quasar intrinsic power (see \S \ref{subsubsec:bolo}), we
obtain quasar bolometric luminosities and Eddington ratios.

\begin{figure}
    \centering
        \includegraphics[width=42mm]{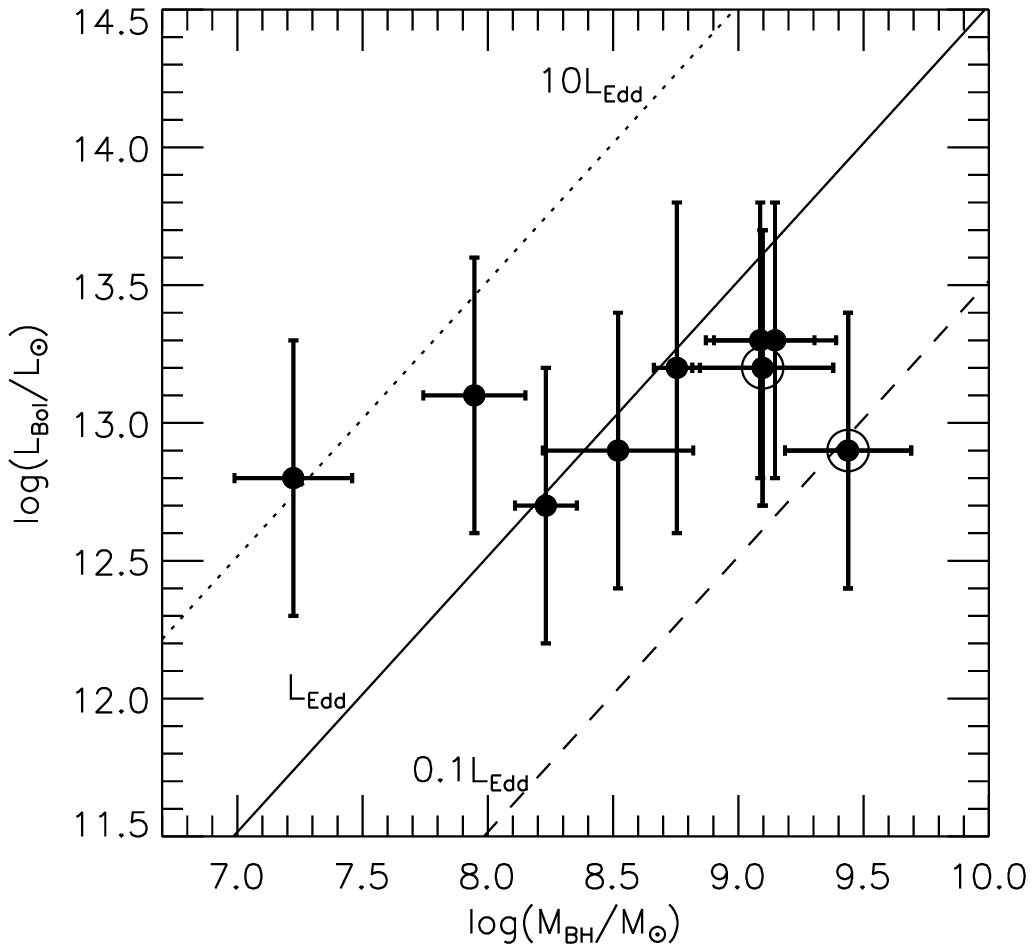}
        \includegraphics[width=42mm]{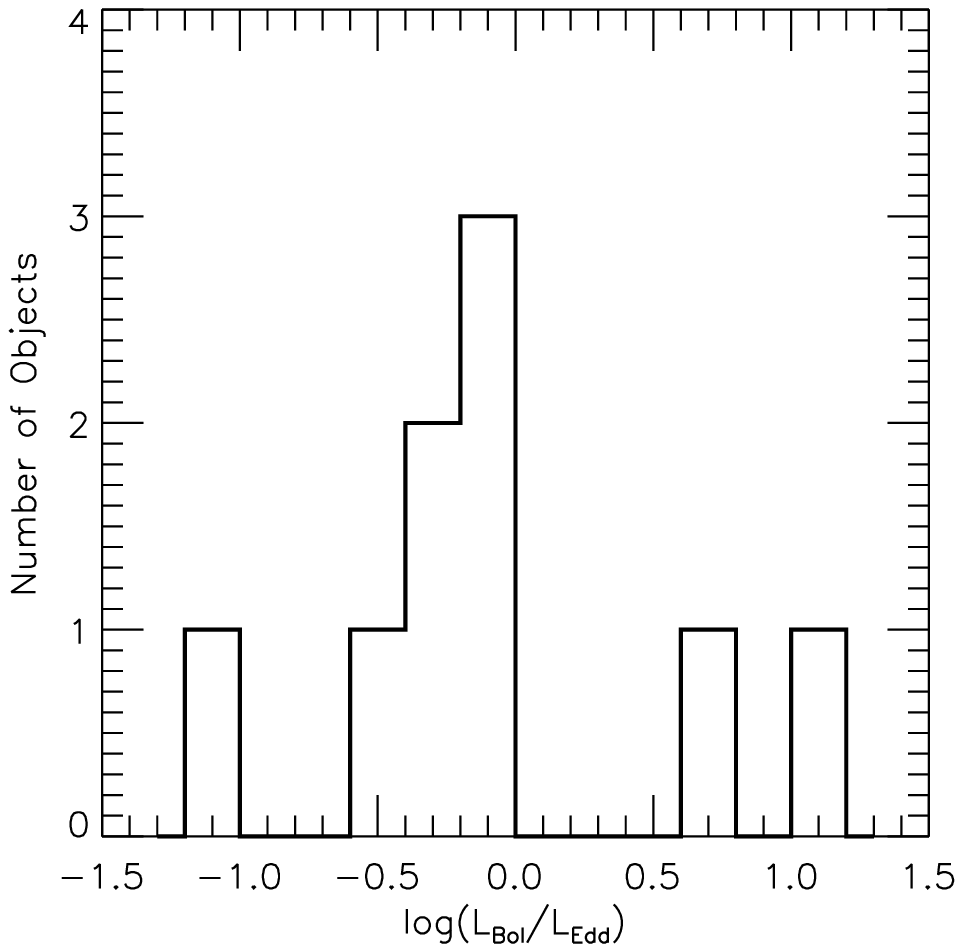}
    \caption{
    Quasar bolometric luminosity $L_{{\rm Bol}}$, black hole mass $M_{{\rm BH}}$, and
    Eddington ratio $L_{{\rm Bol}}/L_{{\rm Edd}}$. {\it Left}: $L_{{\rm Bol}}$ versus $M_{{\rm BH}}$.
    The loci with $L_{{\rm Bol}}/L_{{\rm Edd}}$ equal to 10, 1, and 0.1 are plotted as dotted, solid and dashed lines.
    The two objects indicated with open circles have marginally resolved double cores within the
    central 5 arcsec (\S \ref{subsec:companion}) and the measurements represent luminosity-weighted values of both components.
    {\it Right}: Distribution of Eddington ratios.
    Our sample has a median $L_{{\rm Bol}}/L_{{\rm Edd}}$ of 0.7.
    We list the measurements in Table \ref{tab:sigmafit}.
    See \S \ref{subsec:eddington} for more discussion.
    }
    \label{fig:eddratio}
\end{figure}

Black hole masses, quasar bolometric luminosities, and the
inferred Eddington ratios are listed in Table \ref{tab:sigmafit}
and displayed in Figure \ref{fig:eddratio}. Our sample has a
median black hole mass of $10^{8.8} M_{\odot}$ and a median
Eddington ratio of $\sim 0.7$. As cautioned in \S
\ref{subsec:sigma}, SDSSJ0823$+$3231 and SDSSJ0950$+$0511 have
resolved double cores in their central 5 arcsec which could bias
$\sigma_{\ast}$ towards larger values, resulting in overestimated
black hole masses and underestimated Eddington ratios for the
central nuclei. Taking into account all the associated
uncertainties, almost all of our targets are accreting at higher
than 10\% of the Eddington rates. The two targets estimated to
have super-Eddington ratios both have Wolf-Rayet detections. The
Eddington ratios we find overlap with those of SDSS type 1 quasars
studied by \citet{shen08} with comparable bolometric luminosities
at similar redshifts \citep[also see][]{greene09}.

The contamination on \loiii\, from star formation in our targets
seems negligible given their \oii /\oiii\ emission-line flux
ratios listed in Table \ref{tab:emiratio}
\citep[e.g.,][]{ferland86,ho05}. This statement still holds in the
two super-Eddington targets.

\subsubsection{Bolometric Correction for Type 2 Quasars}\label{subsubsec:bolo}

We estimate bolometric luminosities $L_{{\rm Bol}}$ of type 2
quasars based on \loiii. First, we infer the intrinsic $M_{2500}$
from \loiii\, using the calibration by \citet{reyes08} of the
$M_{2500}$-\loiii\  relation observed in SDSS type 1 quasars. Then
we estimate $L_{{\rm Bol}}$ from $M_{2500}$ using the bolometric
correction for type 1 quasars \citep{marconi04,richards06} and
estimate the systematic uncertainty of the bolometric correction
(the $M_{2500}$ to $L_{{\rm Bol}}$ conversion) to be 0.1-0.2 dex
in the relevant luminosity range (see below). The final
calibration obtained is given by
\begin{equation}\label{eq:bolo}
{\rm log}\bigg(\frac{L_{{\rm Bol}}}{L_{\odot}}\bigg) = 0.99 \times
{\rm log}\bigg(\frac{L_{{\rm [O\,\,  \text{\tiny III}]}}}{L_{\odot}}\bigg) +
3.5,
\end{equation}
with a total 1-$\sigma$ uncertainty of 0.5 dex on log($L_{{\rm
Bol}}/L_{\odot})$ at $L_{{\rm Bol}} \sim 10^{13} L_{\odot}$. The
total uncertainty of $L_{{\rm Bol}}$ convolves the errors
propagated from the measurement error of \loiii, the 0.36 dex
scatter of the $M_{2500}$-\loiii\  relation \citep{reyes08}, the
0.05 dex scatter on $L_B/L_{{\rm Bol}}$ \citep{marconi04}, and the
additional 0.1--0.2 dex systematic uncertainty (depending on
\loiii) from bolometric corrections using different estimates. The
total uncertainty is dominated by the intrinsic scatter of the
$M_{2500}$-\loiii\ relation.

The systematic uncertainty of the bolometric correction for type 1
quasars is estimated by comparing three correction approaches: (1)
converting from $M_{2500}$ to $L_{B}$ assuming $\alpha_{\nu} =
-0.44$ \citep{vandenberk01} and to $L_{{\rm Bol}}$ with the
luminosity-dependent $B$-band bolometric correction of
\citet{marconi04}; (2) converting from $M_{2500}$ to $L_{{\rm
Bol}}$ using the luminosity-independent bolometric correction of
\citet{richards06}; and (3) same as (2), but in the calculation of
the bolometric correction, we extrapolate the 5000--10000
\angstrom\, quasar spectrum into the mid-IR to avoid double
counting of re-radiated emission \citep{marconi04,reyes08}. At
\loiii\,$>10^{9.0} L_{\odot}$, method (2) gives results $<$0.2 dex
higher than those of method (1), while method (3) gives results
$<$0.2 dex lower than those of method (1). We take the results
from method (1) as our baseline values and the differences among
the three as our estimate of the systematic uncertainties.

\subsubsection{Comparing black hole Mass Estimates Based on $\sigma_{\ast}$ and Scattered Quasar
Light}\label{subsubsec:virmass}

For the subset of targets which have broad-H$\beta$ detection (\S
\ref{subsec:scatteredlight}), we could estimate virial black hole
masses using the FWHMs from broad H$\beta$ measurements and the
restored intrinsic quasar continuum luminosity $L_{5100}^{{\rm
QSO}}$ (Table \ref{tab:decomp}) converted from \loiii\, using the
$M_{2500}$-\loiii\  relation of \citet{reyes08} and assuming
$\alpha_{\nu} = -0.44$ \citep{vandenberk01}. Although it depends
on the uncertain H$\beta$ FWHM mass inferences
\citep[e.g.,][]{marconi09}, and on the bolometric corrections for
type 2 quasars, which again have a large systematic uncertainty,
the scattered broad line approach allows us to independently
calibrate the estimation of black hole mass from host galaxy
properties (bulge stellar velocity dispersions and luminosities)
in type 2 quasars.   Results on the black hole mass $M_{{\rm
BH}}^{{\rm vir}}$ using the virial formula
\citep[e.g.][]{greene05} are listed in Table \ref{tab:sigmafit}.
The uncertainty of $M_{{\rm BH}}^{{\rm vir}}$ is dominated by that
propagated from the quasar intrinsic power.   While the two
approaches for estimating black hole mass are based on completely
different assumptions and have separate systematic uncertainties,
the virial estimates based on scattered quasar light are
consistent with $\sigma_{\ast}$-based black hole masses within the
estimated uncertainties.  This general agreement lends further
support to the robustness of both our scattered-light and
$\sigma_{\ast}$ measurements, at least for the small subset of
targets having broad-H$\beta$ detections.   Based on our current
data, we cannot draw any firm conclusion on redshift evolution
considering the large uncertainties and the small sample size.
While the associated uncertainties are large in practice, the
approach offers the possibility to test the $M_{{\rm
BH}}$-$\sigma_{\ast}$ relation in a $\sigma_{\ast}$ range where it
is not well tested at low redshifts \citep{lauer07}.

\subsection{Origins of the Blue Continuum in Luminous Type 2 Quasars}\label{subsec:fc2}

We find that quasar scattered light (\S
\ref{subsec:scatteredlight}) and starlight from young populations
(\S \ref{subsubsec:stellarpop}) both contribute considerably to
the blue continuum observed in luminous type 2 quasars. Their
relative importance varies among different objects, with scattered
light dominating in some cases (e.g., SDSSJ0823$+$3231), and
massive stars dominating in others (e.g., SDSSJ0056$+$0032). In
this section, we first present the dependence of scattered light
on \loiii, and then discuss the implications of the prevalence of
young stellar populations in luminous type 2 quasars.

\subsubsection{A Prescription for Scattered Light}\label{subsubsec:prescription}

The scattered light contamination in low luminosity type 2 AGN
rarely exceeds 5\% in the rest-frame optical
\citep[e.g.][]{cid95,schmitt99,kauffmann03}. However, we find that
it can be significant when \loiii\, is large (\S
\ref{subsec:scatteredlight}), confirming the results of
\citet{zakamska05,zakamska06} based on spectropolarimetry and HST
imaging that a significant fraction of continuum emission can be
due to scattered light in luminous type 2 quasars.  The ratios
$L_{5100}^{{\rm scattered}}/L_{5100}^{{\rm obs}}$ for our nine
targets range from $<5$\% to 55\% (Table \ref{tab:decomp});
combined with the estimated quasar intrinsic luminosity from
\loiii\, (\S \ref{subsubsec:bolo}), the inferred scattered light
fractions $L_{5100}^{{\rm scattered}}/L_{5100}^{{\rm QSO}}$ for
our nine targets range from $<0.2$\% to 0.4\%.   In view of the
high incidence of scattered light in luminous quasars, its
quantification is particularly important for robust determination
of stellar populations (\S \ref{subsec:pop}).

\begin{figure}
    \centering
        \includegraphics[width=70mm]{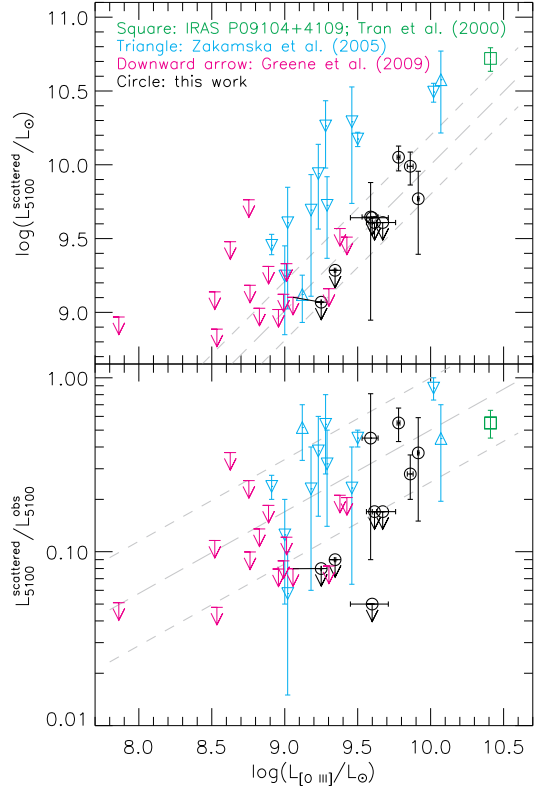}
    \caption{Scattered light contamination as a function of \loiii\,
    in the on-nucleus spectra of luminous type 2 quasars.
    For the nine targets in this work the estimates are based on
    broad-H$\beta$ detections (\S \ref{subsec:scatteredlight})
    and the correlation between H$\beta$ and continuum luminosities
    observed in unobscured quasars \citep{greene05}. Also displayed in comparison
    are scattered-light estimates compiled from the literature including:
    (1) IRAS 09104$+$4109, based on spectropolarimetry from \citet{tran00},
    assuming an intrinsic continuum polarization of 20\%;
    (2) 12 SDSS type 2 quasars, based on continuum decomposition and broad-band
    (downward triangles for yellow band and upward triangles for green band)
    polarization measurements from \citet{zakamska05}, assuming an intrinsic continuum polarization of 20\%;
    and (3) upper limits on scattered light for 15 SDSS type 2 quasars based
    on broad-H$\beta$ measurements \citep{greene09}.
    One of our Gemini targets (SDSSJ0157$-$0053) has also been independently
    studied by \citet{greene09} and its two measurements are connected by a straight line.
    The long dashed line presents the best-fit linear model to all the
    measurements shown, and the dashed lines denote 1-$\sigma$
    scatter. The models are given in Equation \ref{eq:scatter}.
    }
    \label{fig:lo3_vs_sl}
\end{figure}

Figure \ref{fig:lo3_vs_sl} displays both scattered light
luminosity and its fractional contribution as a function of
\loiii. Also shown are measurements compiled from the literature
for comparison (see figure caption for details). The best-fit
linear models to all the data shown are given by
\begin{equation}\label{eq:scatter}
\begin{split}
L_{5100}^{{\rm scattered}} & = 10^{9.5\pm0.3}
\bigg(\frac{L_{{\rm [O \,\, \text{\tiny III}]}}}{10^{9.5}L_{\odot}}\bigg)^{0.99\pm0.02}L_{\odot}, \\
\frac{L_{5100}^{{\rm scattered}}}{L_{5100}^{{\rm obs}}} & =
10^{-0.5\pm0.4} \bigg(\frac{L_{{\rm [O
\,\,\text{\tiny III}]}}}{10^{9.5}L_{\odot}}\bigg)^{0.47\pm0.03}.
\end{split}
\end{equation}

While there is substantial scatter in the above relations and the
amount of scattered light received could vary due to different
coverage of the scattering cone by different observations even at
a given observation angle, the empirical models are useful for a
quick estimate of scattered light given any \loiii .   We caution
that the scattered light fraction estimated here (within the slit)
is a lower bound to the total scattered light fraction, since
there could be considerable amount of scattered light outside our
slit.

\subsubsection{Young Starburst}\label{subsec:starburst}

After the removal of scattered light, there is still considerable
blue light in the spectra of our targets even before reddening
correction. We found significant contributions ($\gtrsim 50$\% of
scattered-light subtracted $L_{5100}$) from very young populations
($\leq$ 25 Myr) in four out of nine of our objects; eight of our
nine targets have $\gtrsim 20$\% of $L_{5100}$ contributions from
very young stellar populations (Table \ref{tab:mass}). In six of
these objects very young stars account for $<5$\% of the stellar
mass.

Considering the short duration of the WR phase (2--5 Myr after a
burst of star formation), the fact that $\gtrsim 30$\% of our
objects show WR features suggests that if the starburst and
Eddington-rate feeding to the SMBH proceed coevally, the
Eddington-rate quasar activity directly associated with the
starburst lasts $\lesssim 10$ Myr, assuming type 2s do not evolve
into type 1s. While a more careful estimate for the timescale can
be obtained by assuming models for the quasar light curve and the
evolving circum-nuclear obscuration, the crude estimate here is in
agreement with that based on black hole demographics
\citep[e.g.][]{yu02}. Enhancing the statistical significance of
our results will require studying a larger sample of
Eddington-rate type 2 quasars. The general approach of using WR
features as a clock demonstrated with our pilot sample here is
another way to observationally estimate the Eddington-rate quasar
duty cycle.

Our data do not have the spatial resolution to determine if these
starburst events are in the immediate circum-nuclear regions or
are occurring on much larger galaxy scales. Nevertheless, it
appears that galaxy interactions are responsible for or coincident
with both the quasar and the starburst activity, at least in the
double-core systems (\S \ref{subsec:companion}).

\subsection{Extinction and Its Effects on Our Results}\label{subsec:dust}

\begin{deluxetable*}{cccccc}
\tabletypesize{\scriptsize}
\tablewidth{0pc}
\tablecaption{
Emission-line Ratios and Color Excess
\label{tab:emiratio}
}
\tablehead{
\colhead{~~~~~~~~~~~~~~~~~~Target Name~~~~~~~~~~~~~~~~~~} &
\colhead{$\frac{F_{{\rm H}\gamma}}{F_{{\rm H}\beta}}$} &
\colhead{$\frac{F_{{\rm H}\delta}}{F_{{\rm H}\beta}}$} &
\colhead{$E(B-V)$} &
\colhead{$E(B-V)$} &
\colhead{$\frac{{\rm [O\,\text{\tiny II}]}}{{\rm [O\,\text{\tiny III}]}}$} \\
\colhead{(1)} &
\colhead{(2)} &
\colhead{(3)} &
\colhead{(4)} &
\colhead{(5)} &
\colhead{(6)}
}
\startdata
SDSSJ0056$+$0032\dotfill & $0.352\pm0.004$ & $0.142\pm0.003$ &  ~~$0.54\pm0.02$ & ~~$0.77\pm0.03$ & $0.14\pm0.01$ \\
SDSSJ0134$+$0014\dotfill & $0.400\pm0.005$ & $0.203\pm0.004$ &  ~~$0.29\pm0.02$ & ~~$0.30\pm0.03$ & $0.09\pm0.01$ \\
SDSSJ0157$-$0053\dotfill & $0.455\pm0.004$ & $0.245\pm0.003$ &  ~~$0.05\pm0.02$ & ~~$0.06\pm0.02$ & $0.15\pm0.01$ \\
SDSSJ0210$-$1001\dotfill & $0.659\pm0.005$ & $0.342\pm0.003$ &   $-0.67\pm0.01$ &  $-0.38\pm0.01$ & $0.17\pm0.01$ \\
SDSSJ0319$-$0058\dotfill & $0.563\pm0.006$ & $0.277\pm0.004$ &   $-0.36\pm0.02$ &  $-0.10\pm0.02$ & $0.14\pm0.05$ \\
SDSSJ0801$+$4412\dotfill & $0.469\pm0.008$ & $0.242\pm0.005$ &   $-0.01\pm0.03$ & ~~$0.07\pm0.03$ & $0.13\pm0.01$ \\
SDSSJ0823$+$3231\dotfill & $0.400\pm0.004$ & $0.224\pm0.003$ &  ~~$0.29\pm0.02$ & ~~$0.18\pm0.02$ & $0.18\pm0.01$ \\
SDSSJ0943$+$3456\dotfill & $0.472\pm0.005$ & $0.199\pm0.003$ &   $-0.02\pm0.02$ & ~~$0.33\pm0.02$ & $0.10\pm0.01$ \\
SDSSJ0950$+$0511\dotfill & $0.383\pm0.007$ & $0.104\pm0.005$ &  ~~$0.38\pm0.04$ & ~~$1.18\pm0.06$ & $0.18\pm0.01$ \\
\enddata
\tablecomments{ \\
Col.(2),(3): Emission-line ratios measured over
continuum-subtracted GMOS spectra
uncorrected for reddening. The quoted 1-$\sigma$ uncertainties were derived based on error spectra.  \\
Col.(4),(5): Color excess estimated from the intensity ratios
$\frac{F_{{\rm H}\gamma}}{F_{{\rm H}\beta}}$ (for Col. 4) and
$\frac{F_{{\rm H}\delta}}{F_{{\rm H}\beta}}$ (for Col. 5), assuming the intrinsic case B values of 0.466
and 0.256 respectively for $T=10^4$ K \citep{osterbrock89} and the extinction
curve of \citet{cardelli89} with $R_V = 3.1$. For most of our
targets (7 out of 9), the two reddening estimates do not agree
with each other and some are negative,
indicating deviations from standard reddening laws and the assumed foreground dust-screen model. \\
Col.(6): Emission-line ratio $\frac{F_{{\rm [O\,
\text{\tiny II}]}\lambda\lambda3727,3729}}{F_{{\rm [O\, \text{\tiny III}]}\lambda5007}}$
characterizing the ionization parameter. The \oii\ $\lambda$3727
line of SDSSJ0319$-$0058 overlaps the GMOS chip gap and its listed
value is measured from SDSS spectra.}
\end{deluxetable*}

Most of the results above are based on spectra uncorrected for
dust extinction in the galaxies themselves. In Table
\ref{tab:emiratio} we list reddening estimates in the narrow-line
regions from the intensity ratios H$\gamma$/H$\beta$ and
H$\delta$/H$\beta$ using the Balmer decrement method, assuming the
intrinsic case B values of 0.466 and 0.256 respectively for
$T=10^4$ K \citep{osterbrock89}, the extinction curve of
\citet{cardelli89} with $R_V = 3.1$, and a model of a foreground
obscuring screen. However, as found by \citet{reyes08} in a subset
of the parent quasar sample having \loiii\,$>$ $10^{9.0}$
$L_{\odot}$, the intensity ratios H$\alpha$/H$\beta$ and
H$\gamma$/H$\beta$ do not obey standard reddening laws, nor can
they be described by a simple dust screen. We find a similar
situation in our sample by comparing the intensity ratios
H$\gamma$/H$\beta$ and H$\delta$/H$\beta$, seven out of nine of
which have significantly different reddening estimates for the
same object (Table \ref{tab:emiratio}).

Despite this caveat, the estimated color excesses $E(B-V)$ for our
9 targets range from $-0.67$ ($-0.38$) to $0.54$ ($1.18$) with a
median of $0.05$ ($0.18$) according to H$\gamma$/H$\beta$
(H$\delta$/H$\beta$). The color excess estimated for the ULIRG,
IRAS 09104$+$4109, by \citet{tran00} is $E(B-V) = 0.24$, also
using the Balmer decrement combining the intensity ratios
H$\gamma$/H$\beta$ and H$\delta$/H$\beta$, a value very similar to
those found in our sample. However, the small color excesses
derived using the Balmer decrement do not necessarily represent
the true dust content of these objects; IRAS 09104$+$4109 is
hyper-luminous in the infrared \citep[$L_{{\rm IR}} \gtrsim
10^{13} L_{\odot}$;][]{kleinmann88} yet has a small color excess
according to the Balmer decrement. This just suggests that most of
the dust is concentrated on scales smaller than the Balmer lines
are emitted. High extinction with little reddening could arise in
objects with very patchy and optically-thick dust clouds.

In view of the uncertainties in extinction geometry \citep[e.g.,
\S 6.2 of][]{reyes08}, we do not apply corrections but rather keep
in mind the potential effects on our results. The unaccounted
extinction would also bias emission-line luminosity measurements
smaller, which means the quasar luminosities and Eddington ratios
(\S \ref{subsec:eddington}; Figure \ref{fig:eddratio}) could be
underestimated. The unaccounted reddening would bias the stellar
ages higher and the estimates of the contribution from young
populations lower (\S \ref{subsubsec:stellarpop}).

\subsection{Comparison with Other Quasar Host Galaxy Studies}\label{subsec:comparison}

\subsubsection{Host Galaxy Studies of Type 2s}\label{subsubsec:type2}

We now compare our results with other host galaxy studies of type
2 AGN. The present work is the first deep optical spectroscopic
study of luminous type 2 quasars with \loiii\,$> 10^{9.3}
L_{\odot}$. A number of previous groups have found that the
presence of young stellar populations is a general property of
type 2 AGN with high \oiii\ luminosities
\citep[e.g.][]{schmitt99,gonzalez01,kauffmann03,cid04,wild07}.
However, the typical ages of young stellar populations ($<$0.1
Gyr) we find are much smaller than the post-starburst ages (1--2
Gyr) seen in the \citet{kauffmann03} subsample with
\loiii$/L_{\odot} \sim 10^{7.0}$--$10^{8.5}$ (reddening
corrected). This result suggests that the luminosity dependence of
host galaxy stellar age and post-starburst fraction seen at lower
luminosities \citep[e.g.][]{heckman80,ho03,kauffmann03} extends to
the most luminous quasars. Within our luminous quasar sample, we
do not see a dependence of stellar age on \loiii, but the scatter
of the correlation (if any) should be large, and our sample may
simply be too small to see such a correlation. A much larger
sample is needed to study the luminosity dependence of host galaxy
properties in luminous type 2 quasars.

Most intriguingly, we detect WR populations in a third of our
sample; WR features are much less often seen at lower luminosities
\citep[e.g.][]{gonzalez01}. The fraction should be treated as a
lower limit since it does not account for regions so obscured that
they do not contribute to the optical emission. If starburst and
quasar activity peak coevally \citep[e.g.][]{dimatteo05}, the
chance of detecting the WR phase would be much larger in luminous
quasars close to the peak activity than in less luminous quasars
which are in their later stages of evolution, after WR features
fade away and intermediate-age post-starburst signatures take
over.

\subsubsection{Host Galaxy Studies of Type 1s}\label{subsubsec:type1}

We now compare our results with those from host galaxy studies of
type 1 quasars and address the issue of whether type 2s and type
1s follow a temporal sequence
\citep[e.g.,][]{sanders88,hopkins06}. Our type 2 targets have
quasar intrinsic luminosities of $M_V < -26$ mag (\S
\ref{subsubsec:bolo}). There is virtually no spectroscopic
information on host galaxies of type 1 quasars at this luminosity
scale \citep[but see][for 3C 48]{boroson82}.

\citet{floyd04} imaged 17 type 1 quasars with $-28 < M_V < -24$
mag and $z \sim 0.4$, and found them to be massive bulge-dominated
galaxies. The Eddington ratios these authors find are comparable
to those of our targets, despite the very different systematic
uncertainties. \citet{urrutia08} studied 13 reddened type 1
quasars with $-26 < M_B < -24$ mag and $z \sim 0.7$, and found
that 11 of them show evidence for recent interactions, a much
higher fraction than that found by \citet{floyd04} and
\citet{guyon06}. While it is possible that deeper high-resolution
imaging will reveal weak interaction features (see below),
\citet{urrutia08} suggest that this inconsistency can be explained
in the evolutionary scenario for quasar obscuration proposed by
\citet{sanders88}, in which dust-reddened type 1 quasars trace an
earlier evolutionary phase than ordinary type 1 quasars, and
thereby are more prone to signs of a merger.

If the evolutionary argument is true, we should also see a high
interaction fraction in our obscured quasar sample.
\citet{zakamska06} found almost half in their obscured quasar
sample show signs of mergers or interactions. Only two of our
targets show double cores (\S \ref{subsec:companion}), but the
stellar population results clearly suggest that our targets trace
an early stage of galaxy interactions if they are responsible for
the young starburst activity.

While the strict AGN unification model states that type 2s are the
same as type 1s other than the difference in our viewing angle
\cite[e.g.,][]{antonucci93}, it is likely that orientation and
evolutionary effects are both at work, and both need to be
considered to interpret the selection of a type 2 quasar sample.
If the opening angle increases as an object proceeds from the
reddened to the ordinary quasar phase and the ordinary phase does
not last much longer than the reddened phase (a statement that
needs to be tested), both of which have $\sim$ Eddington accretion
rates, then by construction (averaged over both orientation and
time), the majority in a sample of luminous type 2 quasars would
be the obscured counterpart of reddened luminous type 1 quasars,
with only a minority resembling regulars. In at least some type 2
quasars, extinction processes are observed to operate both at
small scales due to circum-nuclear dusty tori, and at galaxy
scales due to star formation
\citep[e.g.][]{martinez06,rigby06,lacy07}. Using Spitzer spectra,
\citet{zakamska08} show that SDSS type 2 quasars have the highest
star-formation rate among all quasar samples except for other
samples of type 2 quasars. Luminous type 2 quasars, the majority
of which resemble reddened type 1 quasars, should be sensitive to
the early stages of quasar evolution, and hence are ideal for
quantifying the role played by galaxy interactions.

Quasars with lower luminosities ($-25.5 < M_V < -23.5$ mag) and
with smaller Eddington ratios ($0.01 < L/L_{{Edd}}< 0.1$) almost
all reside in massive ellipticals with relaxed light profiles
\citep{dunlop03}. Deeper imaging studies of a few objects in the
\citet{dunlop03} sample have revealed shells and ripples which are
interpreted as merger relics, with ages of several hundred of Myr
to 1 Gyr estimated based on merger simulations
\citep{canalizo07,bennert08}, consistent with spectroscopically
determined post-starburst ages \citep[e.g.][1.4 Gyr]{canalizo07}.
While shells and ripples are telltale signatures of galaxy
interactions, these systems are observed at a late stage of
mergers, similar to the small subset associated with mergers in
type 2 AGN \citep{kauffmann03}. The typical post-starburst ages of
1--2 Gyr \citep[see also][]{jahnke07,letawe07} are significantly
larger than those of our targets ($<0.1$ Gyr).

In summary, our results combined with those of other quasar host
galaxies spanning different ranges of quasar luminosity and
Eddington ratio may indicate that host galaxy stellar population
properties depend on both quasar luminosity (or more fundamentally
Eddington ratio) and obscuration. This is because the two factors
both characterize the phase of quasar evolution and by extension
that of stellar evolution if the growth of SMBHs and stellar
bulges are coupled. Clearly more spectroscopic data and images
with higher resolution are needed to better understand the host
galaxies of both type 1 and type 2 quasars with luminosities
comparable to luminous $z > 2$ quasars. It is important to
quantify the fraction of objects containing WR populations in
luminous type 1 quasars and to compare with our results for
luminous type 2 quasars, although doing so will be quite
challenging because of the brightness of the quasar continuum.

\section{Summary}\label{sec:sum}

We present deep Gemini GMOS optical spectroscopy of nine luminous
type 2 quasars at redshifts $z \sim 0.5$, drawn from the SDSS type
2 quasar sample of \citet{zakamska03} and \citet{reyes08}. Our
targets were selected to have high intrinsic luminosities ($M_V <
-26$; $L_{{\rm Bol}} \sim 10^{13} L_{\odot}$) inferred by the
\oiii\ luminosity.

Our main findings are as follows:

\begin{enumerate}

\item[1.] We unambiguously detect WR populations in three of our
nine objects. These very young post-starburst populations are
independently confirmed by population modeling of the stellar
continuum: in all three targets with WR detections, a 5-Myr
post-starburst population contributes substantially ($> 50$\% of
$L_{5100}$) to stellar continuum luminosities. In five of the six
targets without WR detections, population modeling shows the
presence of considerable ($> 20$\% of $L_{5100}$) young ($< 0.1$
Gyr) post-starburst populations. We estimate a median SFR of our
sample of $\sim 44$ $M_{\odot}$ yr$^{-1}$ (uncorrected for
extinction).

\item[2.]  We have subtracted continuum scattered light before
performing our stellar population synthesis analysis. The
scattered light is quantified by measuring a broad H$\beta$
component underlying the strong narrow H$\beta$ line. The broad
H$\beta$ component is detected in four targets, and the inferred
scattered light contaminates 30--60\% to the total observed
continuum at 5100 \angstrom. The scattering fraction (within the
slit) is estimated to be $\lesssim 0.4$\% at 5100 \angstrom\, in
all our objects, while we caution it is a lower bound to the total
scattered light fraction, since there could be considerable amount
of scattered light outside our slit. We provide a prescription for
scattered light as a function of \loiii\,(Equation
\ref{eq:scatter}) which can be used to get a quick but rough
estimate of the contamination level.

\item[3.] We obtain stellar velocity dispersion measurements
$\sigma_{\ast}$ for the host galaxies from the G band stellar
absorption feature, which range from 120 to 420 km/s in our
sample. These correspond to black hole masses of $\sim
10^{7.2}$--$10^{9.4} M_{\odot}$ with a median value of $10^{8.8}
M_{\odot}$, assuming the $M_{{\rm BH}}$-$\sigma$ calibration of
\citet{tremaine02}. Combined with the estimated intrinsic quasar
power from \loiii\, (Equation \ref{eq:bolo}), the inferred
Eddington ratios are $\sim$0.1--12, with a median of 0.7 and a
scatter consistent with measurement uncertainties.

\item[4.] Our long-slit spectroscopy shows that four of our nine
targets contain double cores and/or physically associated
companions. At least in three systems, a galaxy interaction might
be responsible for both the quasar and the young starburst
activity, and that luminous type 2 quasars with Eddington ratios
close to unity perhaps trace an early stage of interaction. Our
results combined with those from other quasar host galaxy studies
may suggest that host galaxy stellar population properties depend
on both luminosity and obscuration which characterize the phase of
quasar evolution.

\end{enumerate}

By targeting the obscured population, we are able to study stellar
populations in quasars that are as luminous as the majority of $z
> 2$ quasars. We are directly seeing the youngest stellar
populations in quasar hosts, providing strong evidence for a
direct link between on-going starburst and luminous quasar
activity. However, our pilot sample is too small to study the
correlations between quasar and host galaxy properties, to probe
the physical links and mutual influence of starburst and luminous
quasar activity. Correlation studies of AGN and their hosts at
lower luminosities have yielded extensive physical insight
\citep[e.g.,][]{kauffmann03,heckman04} but are only enabled with
the assembly of statistically large samples. We are going to carry
out more observations in the optical to do such correlation
studies in luminous type 2 quasars at $z \sim 0.5$. We look
forward to the next generation of ground-based near-IR
multi-object spectrographs \citep[e.g.,][]{gunn09} that will
provide statistical samples of luminous obscured quasars at high
redshifts to better understand the coupled growths of stellar
populations and SMBHs in the early universe.

\acknowledgments

We thank R. Reyes, G. Richards, Y. Shen, and P. Smith for helpful
discussions, and D. Norman, J. Gunn, and A. Shapley for the
assistance with our Gemini observations. We thank T. Matheson and
A. Barth for providing the IDL routines for data reduction, and an
anonymous referee for a careful and useful report that improves
the paper.   X.L. and M.A.S. acknowledge the support of NSF grant
AST-0707266.  N.L.Z is supported by the Spitzer Space Telescope
Fellowship provided by NASA through a contract issued by
JPL/Caltech, and by the John N. Bahcall fellowship at the
Institute for Advanced Study. Support for J.E.G. was provided by
NASA through Hubble Fellowship grant HF-01196 awarded by the Space
Telescope Science In- stitute, which is operated by the
Association of Universities for Research in Astronomy, Inc., for
NASA, under contract NAS 5-26555.
Funding for the Sloan Digital Sky Survey (SDSS) has been provided
by the Alfred P. Sloan Foundation, the Participating Institutions,
the National Aeronautics and Space Administration, the National
Science Foundation, the U.S. Department of Energy, the Japanese
Monbukagakusho, and the Max Planck Society. The SDSS Web site is
http://www.sdss.org/.

\bibliography{apj-jour,msigmarefs}

\begin{thebibliography}{102}
\expandafter\ifx\csname natexlab\endcsname\relax\def\natexlab#1{#1}\fi

\bibitem[{{Adelman-McCarthy} {et~al.}(2008){Adelman-McCarthy}, {Ag{\"u}eros},
  {Allam}, {Allende Prieto}, {Anderson}, {Anderson}, {Annis}, {Bahcall},
  {Bailer-Jones}, {Baldry}, {Barentine}, {Bassett}, {Becker}, {Beers}, {Bell},
  {Berlind}, {Bernardi}, {Blanton}, {Bochanski}, {Boroski}, {Brinchmann},
  {Brinkmann}, {Brunner}, {Budav{\'a}ri}, {Carliles}, {Carr}, {Castander},
  {Cinabro}, {Cool}, {Covey}, {Csabai}, {Cunha}, {Davenport}, {Dilday}, {Doi},
  {Eisenstein}, {Evans}, {Fan}, {Finkbeiner}, {Friedman}, {Frieman},
  {Fukugita}, {G{\"a}nsicke}, {Gates}, {Gillespie}, {Glazebrook}, {Gray},
  {Grebel}, {Gunn}, {Gurbani}, {Hall}, {Harding}, {Harvanek}, {Hawley},
  {Hayes}, {Heckman}, {Hendry}, {Hindsley}, {Hirata}, {Hogan}, {Hogg}, {Hyde},
  {Ichikawa}, {Ivezi{\'c}}, {Jester}, {Johnson}, {Jorgensen}, {Juri{\'c}},
  {Kent}, {Kessler}, {Kleinman}, {Knapp}, {Kron}, {Krzesinski}, {Kuropatkin},
  {Lamb}, {Lampeitl}, {Lebedeva}, {Lee}, {Leger}, {L{\'e}pine}, {Lima}, {Lin},
  {Long}, {Loomis}, {Loveday}, {Lupton}, {Malanushenko}, {Malanushenko},
  {Mandelbaum}, {Margon}, {Marriner}, {Mart{\'{\i}}nez-Delgado}, {Matsubara},
  {McGehee}, {McKay}, {Meiksin}, {Morrison}, {Munn}, {Nakajima}, {Neilsen},
  {Newberg}, {Nichol}, {Nicinski}, {Nieto-Santisteban}, {Nitta}, {Okamura},
  {Owen}, {Oyaizu}, {Padmanabhan}, {Pan}, {Park}, {Peoples}, {Pier}, {Pope},
  {Purger}, {Raddick}, {Re Fiorentin}, {Richards}, {Richmond}, {Riess}, {Rix},
  {Rockosi}, {Sako}, {Schlegel}, {Schneider}, {Schreiber}, {Schwope}, {Seljak},
  {Sesar}, {Sheldon}, {Shimasaku}, {Sivarani}, {Smith}, {Snedden}, {Steinmetz},
  {Strauss}, {SubbaRao}, {Suto}, {Szalay}, {Szapudi}, {Szkody}, {Tegmark},
  {Thakar}, {Tremonti}, {Tucker}, {Uomoto}, {Vanden Berk}, {Vandenberg},
  {Vidrih}, {Vogeley}, {Voges}, {Vogt}, {Wadadekar}, {Weinberg}, {West},
  {White}, {Wilhite}, {Yanny}, {Yocum}, {York}, {Zehavi}, \& {Zucker}}]{dr6}
{Adelman-McCarthy}, J.~K., {et~al.} 2008, \apjs, 175, 297

\bibitem[{{Antonucci}(1993)}]{antonucci93}
{Antonucci}, R. 1993, \araa, 31, 473

\bibitem[{{Armus} {et~al.}(1988){Armus}, {Heckman}, \& {Miley}}]{armus88}
{Armus}, L., {Heckman}, T.~M., \& {Miley}, G.~K. 1988, \apjl, 326, L45

\bibitem[{{Bahcall} {et~al.}(1997){Bahcall}, {Kirhakos}, {Saxe}, \&
  {Schneider}}]{bahcall97}
{Bahcall}, J.~N., {Kirhakos}, S., {Saxe}, D.~H., \& {Schneider}, D.~P. 1997,
  \apj, 479, 642

\bibitem[{{Bell} {et~al.}(2009){Bell}, {Davis}, {Dey}, {van Dokkum}, {Ellis},
  {Eisenstein}, {Elvis}, {Faber}, {Frenk}, {Genzel}, {Greene}, {Gunn},
  {Kauffmann}, {Knapp}, {Kriek}, {Larkin}, {Maraston}, {Nandra}, {Ostriker},
  {Prada}, {Schlegel}, {Strauss}, {Szalay}, {Tremonti}, {White}, {White}, \&
  {Wyse}}]{gunn09}
{Bell}, E., {et~al.} 2009, ArXiv e-prints

\bibitem[{{Bennert} {et~al.}(2008){Bennert}, {Canalizo}, {Jungwiert},
  {Stockton}, {Schweizer}, {Peng}, \& {Lacy}}]{bennert08}
{Bennert}, N., {Canalizo}, G., {Jungwiert}, B., {Stockton}, A., {Schweizer},
  F., {Peng}, C.~Y., \& {Lacy}, M. 2008, \apj, 677, 846

\bibitem[{{Bernardi} {et~al.}(2006){Bernardi}, {Sheth}, {Nichol}, {Miller},
  {Schlegel}, {Frieman}, {Schneider}, {Subbarao}, {York}, \&
  {Brinkmann}}]{bernardi06}
{Bernardi}, M., {et~al.} 2006, \aj, 131, 2018

\bibitem[{{Boroson} \& {Oke}(1982)}]{boroson82}
{Boroson}, T.~A., \& {Oke}, J.~B. 1982, \nat, 296, 397

\bibitem[{{Boyle} {et~al.}(2000){Boyle}, {Shanks}, {Croom}, {Smith}, {Miller},
  {Loaring}, \& {Heymans}}]{boyle00}
{Boyle}, B.~J., {Shanks}, T., {Croom}, S.~M., {Smith}, R.~J., {Miller}, L.,
  {Loaring}, N., \& {Heymans}, C. 2000, \mnras, 317, 1014

\bibitem[{{Boyle} \& {Terlevich}(1998)}]{boyle98}
{Boyle}, B.~J., \& {Terlevich}, R.~J. 1998, \mnras, 293, L49

\bibitem[{{Brinchmann} {et~al.}(2008){Brinchmann}, {Kunth}, \&
  {Durret}}]{brinchmann08}
{Brinchmann}, J., {Kunth}, D., \& {Durret}, F. 2008, \aap, 485, 657

\bibitem[{{Bruzual} \& {Charlot}(2003)}]{bc03}
{Bruzual}, G., \& {Charlot}, S. 2003, \mnras, 344, 1000

\bibitem[{{Canalizo} {et~al.}(2007){Canalizo}, {Bennert}, {Jungwiert},
  {Stockton}, {Schweizer}, {Lacy}, \& {Peng}}]{canalizo07}
{Canalizo}, G., {Bennert}, N., {Jungwiert}, B., {Stockton}, A., {Schweizer},
  F., {Lacy}, M., \& {Peng}, C. 2007, \apj, 669, 801

\bibitem[{{Canalizo} \& {Stockton}(2000)}]{canalizo00}
{Canalizo}, G., \& {Stockton}, A. 2000, \apj, 528, 201

\bibitem[{{Canalizo} {et~al.}(2008){Canalizo}, {Wold}, {Lazarova}, \&
  {Lacy}}]{canalizo08}
{Canalizo}, G., {Wold}, M., {Lazarova}, M., \& {Lacy}, M. 2008, in American
  Institute of Physics Conference Series, Vol. 1053, American Institute of
  Physics Conference Series, 63--66

\bibitem[{{Cardelli} {et~al.}(1989){Cardelli}, {Clayton}, \&
  {Mathis}}]{cardelli89}
{Cardelli}, J.~A., {Clayton}, G.~C., \& {Mathis}, J.~S. 1989, \apj, 345, 245

\bibitem[{{Chabrier}(2003)}]{chabrier03}
{Chabrier}, G. 2003, \pasp, 115, 763

\bibitem[{{Chapman} {et~al.}(2005){Chapman}, {Blain}, {Smail}, \&
  {Ivison}}]{chapman05}
{Chapman}, S.~C., {Blain}, A.~W., {Smail}, I., \& {Ivison}, R.~J. 2005, \apj,
  622, 772

\bibitem[{{Cid Fernandes} {et~al.}(2004){Cid Fernandes}, {Gu}, {Melnick},
  {Terlevich}, {Terlevich}, {Kunth}, {Rodrigues Lacerda}, \& {Joguet}}]{cid04}
{Cid Fernandes}, R., {Gu}, Q., {Melnick}, J., {Terlevich}, E., {Terlevich}, R.,
  {Kunth}, D., {Rodrigues Lacerda}, R., \& {Joguet}, B. 2004, \mnras, 355, 273

\bibitem[{{Cid Fernandes} \& {Terlevich}(1995)}]{cid95}
{Cid Fernandes}, R.~J., \& {Terlevich}, R. 1995, \mnras, 272, 423

\bibitem[{{Connolly} {et~al.}(1997){Connolly}, {Szalay}, {Dickinson},
  {Subbarao}, \& {Brunner}}]{connolly97}
{Connolly}, A.~J., {Szalay}, A.~S., {Dickinson}, M., {Subbarao}, M.~U., \&
  {Brunner}, R.~J. 1997, \apjl, 486, L11+

\bibitem[{{De Robertis} {et~al.}(1998){De Robertis}, {Yee}, \&
  {Hayhoe}}]{derobertis98}
{De Robertis}, M.~M., {Yee}, H.~K.~C., \& {Hayhoe}, K. 1998, \apj, 496, 93

\bibitem[{{Di Matteo} {et~al.}(2005){Di Matteo}, {Springel}, \&
  {Hernquist}}]{dimatteo05}
{Di Matteo}, T., {Springel}, V., \& {Hernquist}, L. 2005, \nat, 433, 604

\bibitem[{{Dunlop} {et~al.}(2003){Dunlop}, {McLure}, {Kukula}, {Baum}, {O'Dea},
  \& {Hughes}}]{dunlop03}
{Dunlop}, J.~S., {McLure}, R.~J., {Kukula}, M.~J., {Baum}, S.~A., {O'Dea},
  C.~P., \& {Hughes}, D.~H. 2003, \mnras, 340, 1095

\bibitem[{{Ellison} {et~al.}(2008){Ellison}, {Patton}, {Simard}, \&
  {McConnachie}}]{ellison08}
{Ellison}, S.~L., {Patton}, D.~R., {Simard}, L., \& {McConnachie}, A.~W. 2008,
  \aj, 135, 1877

\bibitem[{{Fabian}(1999)}]{fabian99}
{Fabian}, A.~C. 1999, \mnras, 308, L39

\bibitem[{{Ferland} \& {Osterbrock}(1986)}]{ferland86}
{Ferland}, G.~J., \& {Osterbrock}, D.~E. 1986, \apj, 300, 658

\bibitem[{{Ferrarese} \& {Merritt}(2000)}]{ferrarese00}
{Ferrarese}, L., \& {Merritt}, D. 2000, \apjl, 539, L9

\bibitem[{{Floyd} {et~al.}(2004){Floyd}, {Kukula}, {Dunlop}, {McLure},
  {Miller}, {Percival}, {Baum}, \& {O'Dea}}]{floyd04}
{Floyd}, D.~J.~E., {Kukula}, M.~J., {Dunlop}, J.~S., {McLure}, R.~J., {Miller},
  L., {Percival}, W.~J., {Baum}, S.~A., \& {O'Dea}, C.~P. 2004, \mnras, 355,
  196

\bibitem[{{Fuentes-Williams} \& {Stocke}(1988)}]{fuentes88}
{Fuentes-Williams}, T., \& {Stocke}, J.~T. 1988, \aj, 96, 1235

\bibitem[{{Gebhardt} {et~al.}(2000){Gebhardt}, {Bender}, {Bower}, {Dressler},
  {Faber}, {Filippenko}, {Green}, {Grillmair}, {Ho}, {Kormendy}, {Lauer},
  {Magorrian}, {Pinkney}, {Richstone}, \& {Tremaine}}]{gebhardt00}
{Gebhardt}, K., {et~al.} 2000, \apjl, 539, L13

\bibitem[{{Gonz{\'a}lez Delgado} {et~al.}(2001){Gonz{\'a}lez Delgado},
  {Heckman}, \& {Leitherer}}]{gonzalez01}
{Gonz{\'a}lez Delgado}, R.~M., {Heckman}, T., \& {Leitherer}, C. 2001, \apj,
  546, 845

\bibitem[{{Goodrich} \& {Miller}(1989)}]{goodrich89}
{Goodrich}, R.~W., \& {Miller}, J.~S. 1989, \apjl, 346, L21

\bibitem[{{Greene} \& {Ho}(2005)}]{greene05}
{Greene}, J.~E., \& {Ho}, L.~C. 2005, \apj, 630, 122

\bibitem[{{Greene} \& {Ho}(2006{\natexlab{a}})}]{greene06a}
---. 2006{\natexlab{a}}, \apj, 641, 117

\bibitem[{{Greene} \& {Ho}(2006{\natexlab{b}})}]{greene06b}
---. 2006{\natexlab{b}}, \apjl, 641, L21

\bibitem[{{Greene} {et~al.}(2009){Greene}, {Zakamska}, {Liu}, {Barth}, \&
  {Ho}}]{greene09}
{Greene}, J.~E., {Zakamska}, N.~L., {Liu}, X., {Barth}, A.~J., \& {Ho}, L.~C.
  2009, ArXiv e-prints

\bibitem[{{Guyon} {et~al.}(2006){Guyon}, {Sanders}, \& {Stockton}}]{guyon06}
{Guyon}, O., {Sanders}, D.~B., \& {Stockton}, A. 2006, \apjs, 166, 89

\bibitem[{{Hasinger} {et~al.}(2005){Hasinger}, {Miyaji}, \&
  {Schmidt}}]{hasinger05}
{Hasinger}, G., {Miyaji}, T., \& {Schmidt}, M. 2005, \aap, 441, 417

\bibitem[{{Heckman} {et~al.}(1995){Heckman}, {Krolik}, {Meurer}, {Calzetti},
  {Kinney}, {Koratkar}, {Leitherer}, {Robert}, \& {Wilson}}]{heckman95}
{Heckman}, T., {et~al.} 1995, \apj, 452, 549

\bibitem[{{Heckman}(1980)}]{heckman80}
{Heckman}, T.~M. 1980, \aap, 87, 142

\bibitem[{{Heckman} {et~al.}(1997){Heckman}, {Gonzalez-Delgado}, {Leitherer},
  {Meurer}, {Krolik}, {Wilson}, {Koratkar}, \& {Kinney}}]{heckman97}
{Heckman}, T.~M., {Gonzalez-Delgado}, R., {Leitherer}, C., {Meurer}, G.~R.,
  {Krolik}, J., {Wilson}, A.~S., {Koratkar}, A., \& {Kinney}, A. 1997, \apj,
  482, 114

\bibitem[{{Heckman} {et~al.}(2004){Heckman}, {Kauffmann}, {Brinchmann},
  {Charlot}, {Tremonti}, \& {White}}]{heckman04}
{Heckman}, T.~M., {Kauffmann}, G., {Brinchmann}, J., {Charlot}, S., {Tremonti},
  C., \& {White}, S.~D.~M. 2004, \apj, 613, 109

\bibitem[{{Ho}(2005)}]{ho05}
{Ho}, L.~C. 2005, \apj, 629, 680

\bibitem[{{Ho} {et~al.}(2008){Ho}, {Darling}, \& {Greene}}]{ho08b}
{Ho}, L.~C., {Darling}, J., \& {Greene}, J.~E. 2008, \apj, 681, 128

\bibitem[{{Ho} {et~al.}(2003){Ho}, {Filippenko}, \& {Sargent}}]{ho03}
{Ho}, L.~C., {Filippenko}, A.~V., \& {Sargent}, W.~L.~W. 2003, \apj, 583, 159

\bibitem[{{Ho} {et~al.}(2009){Ho}, {Greene}, {Filippenko}, \& {Sargent}}]{ho09}
{Ho}, L.~C., {Greene}, J.~E., {Filippenko}, A.~V., \& {Sargent}, W.~L.~W. 2009,
  \apjs, 183, 1

\bibitem[{{Hopkins} {et~al.}(2006){Hopkins}, {Hernquist}, {Cox}, {Di Matteo},
  {Robertson}, \& {Springel}}]{hopkins06}
{Hopkins}, P.~F., {Hernquist}, L., {Cox}, T.~J., {Di Matteo}, T., {Robertson},
  B., \& {Springel}, V. 2006, \apjs, 163, 1

\bibitem[{{Hutchings} \& {Crampton}(1990)}]{hutchings90}
{Hutchings}, J.~B., \& {Crampton}, D. 1990, \aj, 99, 37

\bibitem[{{Jahnke} {et~al.}(2004){Jahnke}, {Kuhlbrodt}, \&
  {Wisotzki}}]{jahnke04}
{Jahnke}, K., {Kuhlbrodt}, B., \& {Wisotzki}, L. 2004, \mnras, 352, 399

\bibitem[{{Jahnke} {et~al.}(2007){Jahnke}, {Wisotzki}, {Courbin}, \&
  {Letawe}}]{jahnke07}
{Jahnke}, K., {Wisotzki}, L., {Courbin}, F., \& {Letawe}, G. 2007, \mnras, 378,
  23

\bibitem[{{Kauffmann} {et~al.}(2003){Kauffmann}, {Heckman}, {Tremonti},
  {Brinchmann}, {Charlot}, {White}, {Ridgway}, {Brinkmann}, {Fukugita}, {Hall},
  {Ivezi{\'c}}, {Richards}, \& {Schneider}}]{kauffmann03}
{Kauffmann}, G., {et~al.} 2003, \mnras, 346, 1055

\bibitem[{{Kay}(1994)}]{kay94}
{Kay}, L.~E. 1994, \apj, 430, 196

\bibitem[{{Kennicutt}(1998)}]{kennicutt98}
{Kennicutt}, Jr., R.~C. 1998, \araa, 36, 189

\bibitem[{{Kirhakos} {et~al.}(1999){Kirhakos}, {Bahcall}, {Schneider}, \&
  {Kristian}}]{kirhakos99}
{Kirhakos}, S., {Bahcall}, J.~N., {Schneider}, D.~P., \& {Kristian}, J. 1999,
  \apj, 520, 67

\bibitem[{{Kleinmann} {et~al.}(1988){Kleinmann}, {Hamilton}, {Keel},
  {Wynn-Williams}, {Eales}, {Becklin}, \& {Kuntz}}]{kleinmann88}
{Kleinmann}, S.~G., {Hamilton}, D., {Keel}, W.~C., {Wynn-Williams}, C.~G.,
  {Eales}, S.~A., {Becklin}, E.~E., \& {Kuntz}, K.~D. 1988, \apj, 328, 161

\bibitem[{{Kormendy} \& {Richstone}(1995)}]{kormendy95}
{Kormendy}, J., \& {Richstone}, D. 1995, \araa, 33, 581

\bibitem[{{Kunth} \& {Contini}(1999)}]{kunth99}
{Kunth}, D., \& {Contini}, T. 1999, in IAU Symposium, Vol. 193, Wolf-Rayet
  Phenomena in Massive Stars and Starburst Galaxies, ed. K.~A. {van der Hucht},
  G.~{Koenigsberger}, \& P.~R.~J. {Eenens}, 725--+

\bibitem[{{Lacy} {et~al.}(2007){Lacy}, {Sajina}, {Petric}, {Seymour},
  {Canalizo}, {Ridgway}, {Armus}, \& {Storrie-Lombardi}}]{lacy07}
{Lacy}, M., {Sajina}, A., {Petric}, A.~O., {Seymour}, N., {Canalizo}, G.,
  {Ridgway}, S.~E., {Armus}, L., \& {Storrie-Lombardi}, L.~J. 2007, \apjl, 669,
  L61

\bibitem[{{Lauer} {et~al.}(2007){Lauer}, {Tremaine}, {Richstone}, \&
  {Faber}}]{lauer07}
{Lauer}, T.~R., {Tremaine}, S., {Richstone}, D., \& {Faber}, S.~M. 2007, \apj,
  670, 249

\bibitem[{{Letawe} {et~al.}(2007){Letawe}, {Magain}, {Courbin}, {Jablonka},
  {Jahnke}, {Meylan}, \& {Wisotzki}}]{letawe07}
{Letawe}, G., {Magain}, P., {Courbin}, F., {Jablonka}, P., {Jahnke}, K.,
  {Meylan}, G., \& {Wisotzki}, L. 2007, \mnras, 378, 83

\bibitem[{{Li} {et~al.}(2008){Li}, {Kauffmann}, {Heckman}, {White}, \&
  {Jing}}]{li08}
{Li}, C., {Kauffmann}, G., {Heckman}, T.~M., {White}, S.~D.~M., \& {Jing},
  Y.~P. 2008, \mnras, 385, 1915

\bibitem[{{Magain} {et~al.}(2005){Magain}, {Letawe}, {Courbin}, {Jablonka},
  {Jahnke}, {Meylan}, \& {Wisotzki}}]{magain05}
{Magain}, P., {Letawe}, G., {Courbin}, F., {Jablonka}, P., {Jahnke}, K.,
  {Meylan}, G., \& {Wisotzki}, L. 2005, \nat, 437, 381

\bibitem[{{Magorrian} {et~al.}(1998){Magorrian}, {Tremaine}, {Richstone},
  {Bender}, {Bower}, {Dressler}, {Faber}, {Gebhardt}, {Green}, {Grillmair},
  {Kormendy}, \& {Lauer}}]{magorrian98}
{Magorrian}, J., {et~al.} 1998, \aj, 115, 2285

\bibitem[{{Marconi} {et~al.}(2009){Marconi}, {Axon}, {Maiolino}, {Nagao},
  {Pietrini}, {Risaliti}, {Robinson}, \& {Torricelli}}]{marconi09}
{Marconi}, A., {Axon}, D.~J., {Maiolino}, R., {Nagao}, T., {Pietrini}, P.,
  {Risaliti}, G., {Robinson}, A., \& {Torricelli}, G. 2009, \apjl, 698, L103

\bibitem[{{Marconi} {et~al.}(2004){Marconi}, {Risaliti}, {Gilli}, {Hunt},
  {Maiolino}, \& {Salvati}}]{marconi04}
{Marconi}, A., {Risaliti}, G., {Gilli}, R., {Hunt}, L.~K., {Maiolino}, R., \&
  {Salvati}, M. 2004, \mnras, 351, 169

\bibitem[{{Mart{\'{\i}}nez-Sansigre} {et~al.}(2006){Mart{\'{\i}}nez-Sansigre},
  {Rawlings}, {Lacy}, {Fadda}, {Jarvis}, {Marleau}, {Simpson}, \&
  {Willott}}]{martinez06}
{Mart{\'{\i}}nez-Sansigre}, A., {Rawlings}, S., {Lacy}, M., {Fadda}, D.,
  {Jarvis}, M.~J., {Marleau}, F.~R., {Simpson}, C., \& {Willott}, C.~J. 2006,
  \mnras, 370, 1479

\bibitem[{{Matheson} {et~al.}(2008){Matheson}, {Kirshner}, {Challis}, {Jha},
  {Garnavich}, {Berlind}, {Calkins}, {Blondin}, {Balog}, {Bragg}, {Caldwell},
  {Dendy Concannon}, {Falco}, {Graves}, {Huchra}, {Kuraszkiewicz}, {Mader},
  {Mahdavi}, {Phelps}, {Rines}, {Song}, \& {Wilkes}}]{matheson08}
{Matheson}, T., {et~al.} 2008, \aj, 135, 1598

\bibitem[{{McLure} {et~al.}(1999){McLure}, {Kukula}, {Dunlop}, {Baum}, {O'Dea},
  \& {Hughes}}]{mclure99}
{McLure}, R.~J., {Kukula}, M.~J., {Dunlop}, J.~S., {Baum}, S.~A., {O'Dea},
  C.~P., \& {Hughes}, D.~H. 1999, \mnras, 308, 377

\bibitem[{{Nolan} {et~al.}(2001){Nolan}, {Dunlop}, {Kukula}, {Hughes},
  {Boroson}, \& {Jimenez}}]{nolan01}
{Nolan}, L.~A., {Dunlop}, J.~S., {Kukula}, M.~J., {Hughes}, D.~H., {Boroson},
  T., \& {Jimenez}, R. 2001, \mnras, 323, 308

\bibitem[{{Osterbrock}(1989)}]{osterbrock89}
{Osterbrock}, D.~E. 1989, {Astrophysics of gaseous nebulae and active galactic
  nuclei} (Research supported by the University of California, John Simon
  Guggenheim Memorial Foundation, University of Minnesota, et al.~Mill Valley,
  CA, University Science Books, 1989, 422 p.)

\bibitem[{{Osterbrock} \& {Cohen}(1982)}]{osterbrock82}
{Osterbrock}, D.~E., \& {Cohen}, R.~D. 1982, \apj, 261, 64

\bibitem[{{Pierce} {et~al.}(2007){Pierce}, {Lotz}, {Laird}, {Lin}, {Nandra},
  {Primack}, {Faber}, {Barmby}, {Park}, {Willner}, {Gwyn}, {Koo}, {Coil},
  {Cooper}, {Georgakakis}, {Koekemoer}, {Noeske}, {Weiner}, \&
  {Willmer}}]{pierce07}
{Pierce}, C.~M., {et~al.} 2007, \apjl, 660, L19

\bibitem[{{Reichard} {et~al.}(2009){Reichard}, {Heckman}, {Rudnick},
  {Brinchmann}, {Kauffmann}, \& {Wild}}]{reichard09}
{Reichard}, T.~A., {Heckman}, T.~M., {Rudnick}, G., {Brinchmann}, J.,
  {Kauffmann}, G., \& {Wild}, V. 2009, \apj, 691, 1005

\bibitem[{{Reyes} {et~al.}(2008){Reyes}, {Zakamska}, {Strauss}, {Green},
  {Krolik}, {Shen}, {Richards}, {Anderson}, \& {Schneider}}]{reyes08}
{Reyes}, R., {et~al.} 2008, \aj, 136, 2373

\bibitem[{{Richards} {et~al.}(2006){Richards}, {Strauss}, {Fan}, {Hall},
  {Jester}, {Schneider}, {Vanden Berk}, {Stoughton}, {Anderson}, {Brunner},
  {Gray}, {Gunn}, {Ivezi{\'c}}, {Kirkland}, {Knapp}, {Loveday}, {Meiksin},
  {Pope}, {Szalay}, {Thakar}, {Yanny}, {York}, {Barentine}, {Brewington},
  {Brinkmann}, {Fukugita}, {Harvanek}, {Kent}, {Kleinman}, {Krzesi{\'n}ski},
  {Long}, {Lupton}, {Nash}, {Neilsen}, {Nitta}, {Schlegel}, \&
  {Snedden}}]{richards06}
{Richards}, G.~T., {et~al.} 2006, \aj, 131, 2766

\bibitem[{{Rigby} {et~al.}(2006){Rigby}, {Rieke}, {Donley}, {Alonso-Herrero},
  \& {P{\'e}rez-Gonz{\'a}lez}}]{rigby06}
{Rigby}, J.~R., {Rieke}, G.~H., {Donley}, J.~L., {Alonso-Herrero}, A., \&
  {P{\'e}rez-Gonz{\'a}lez}, P.~G. 2006, \apj, 645, 115

\bibitem[{{Sanders} {et~al.}(1988){Sanders}, {Soifer}, {Elias}, {Madore},
  {Matthews}, {Neugebauer}, \& {Scoville}}]{sanders88}
{Sanders}, D.~B., {Soifer}, B.~T., {Elias}, J.~H., {Madore}, B.~F., {Matthews},
  K., {Neugebauer}, G., \& {Scoville}, N.~Z. 1988, \apj, 325, 74

\bibitem[{{Schaerer} \& {Vacca}(1998)}]{schaerer98}
{Schaerer}, D., \& {Vacca}, W.~D. 1998, \apj, 497, 618

\bibitem[{{Schmitt} {et~al.}(1999){Schmitt}, {Storchi-Bergmann}, \&
  {Fernandes}}]{schmitt99}
{Schmitt}, H.~R., {Storchi-Bergmann}, T., \& {Fernandes}, R.~C. 1999, \mnras,
  303, 173

\bibitem[{{Shen}(2009)}]{shen09}
{Shen}, Y. 2009, ArXiv e-prints

\bibitem[{{Shen} {et~al.}(2008){Shen}, {Greene}, {Strauss}, {Richards}, \&
  {Schneider}}]{shen08}
{Shen}, Y., {Greene}, J.~E., {Strauss}, M.~A., {Richards}, G.~T., \&
  {Schneider}, D.~P. 2008, \apj, 680, 169

\bibitem[{{Silk} \& {Rees}(1998)}]{silk98}
{Silk}, J., \& {Rees}, M.~J. 1998, \aap, 331, L1

\bibitem[{{Smith}(1991)}]{smith91}
{Smith}, L.~F. 1991, in IAU Symposium, Vol. 143, Wolf-Rayet Stars and
  Interrelations with Other Massive Stars in Galaxies, ed. K.~A. {van der
  Hucht} \& B.~{Hidayat}, 601--612

\bibitem[{{Smith} {et~al.}(1996){Smith}, {Shara}, \& {Moffat}}]{smith96}
{Smith}, L.~F., {Shara}, M.~M., \& {Moffat}, A.~F.~J. 1996, \mnras, 281, 163

\bibitem[{{Tran}(1995)}]{tran95}
{Tran}, H.~D. 1995, \apj, 440, 597

\bibitem[{{Tran} {et~al.}(2000){Tran}, {Cohen}, \& {Villar-Martin}}]{tran00}
{Tran}, H.~D., {Cohen}, M.~H., \& {Villar-Martin}, M. 2000, \aj, 120, 562

\bibitem[{{Tremaine} {et~al.}(2002){Tremaine}, {Gebhardt}, {Bender}, {Bower},
  {Dressler}, {Faber}, {Filippenko}, {Green}, {Grillmair}, {Ho}, {Kormendy},
  {Lauer}, {Magorrian}, {Pinkney}, \& {Richstone}}]{tremaine02}
{Tremaine}, S., {et~al.} 2002, \apj, 574, 740

\bibitem[{{Urrutia} {et~al.}(2008){Urrutia}, {Lacy}, \& {Becker}}]{urrutia08}
{Urrutia}, T., {Lacy}, M., \& {Becker}, R.~H. 2008, \apj, 674, 80

\bibitem[{{Vacca} \& {Conti}(1992)}]{vacca92}
{Vacca}, W.~D., \& {Conti}, P.~S. 1992, \apj, 401, 543

\bibitem[{{Valdes} {et~al.}(2004){Valdes}, {Gupta}, {Rose}, {Singh}, \&
  {Bell}}]{valdes04}
{Valdes}, F., {Gupta}, R., {Rose}, J.~A., {Singh}, H.~P., \& {Bell}, D.~J.
  2004, \apjs, 152, 251

\bibitem[{{Vanden Berk} {et~al.}(2001){Vanden Berk}, {Richards}, {Bauer},
  {Strauss}, {Schneider}, {Heckman}, {York}, {Hall}, {Fan}, {Knapp},
  {Anderson}, {Annis}, {Bahcall}, {Bernardi}, {Briggs}, {Brinkmann}, {Brunner},
  {Burles}, {Carey}, {Castander}, {Connolly}, {Crocker}, {Csabai}, {Doi},
  {Finkbeiner}, {Friedman}, {Frieman}, {Fukugita}, {Gunn}, {Hennessy},
  {Ivezi{\'c}}, {Kent}, {Kunszt}, {Lamb}, {Leger}, {Long}, {Loveday}, {Lupton},
  {Meiksin}, {Merelli}, {Munn}, {Newberg}, {Newcomb}, {Nichol}, {Owen}, {Pier},
  {Pope}, {Rockosi}, {Schlegel}, {Siegmund}, {Smee}, {Snir}, {Stoughton},
  {Stubbs}, {SubbaRao}, {Szalay}, {Szokoly}, {Tremonti}, {Uomoto}, {Waddell},
  {Yanny}, \& {Zheng}}]{vandenberk01}
{Vanden Berk}, D.~E., {et~al.} 2001, \aj, 122, 549

\bibitem[{{Watson} {et~al.}(2008){Watson}, {Martini}, {Dasyra}, {Bentz},
  {Ferrarese}, {Peterson}, {Pogge}, \& {Tacconi}}]{watson08}
{Watson}, L.~C., {Martini}, P., {Dasyra}, K.~M., {Bentz}, M.~C., {Ferrarese},
  L., {Peterson}, B.~M., {Pogge}, R.~W., \& {Tacconi}, L.~J. 2008, \apjl, 682,
  L21

\bibitem[{{Wild} {et~al.}(2007){Wild}, {Kauffmann}, {Heckman}, {Charlot},
  {Lemson}, {Brinchmann}, {Reichard}, \& {Pasquali}}]{wild07}
{Wild}, V., {Kauffmann}, G., {Heckman}, T., {Charlot}, S., {Lemson}, G.,
  {Brinchmann}, J., {Reichard}, T., \& {Pasquali}, A. 2007, \mnras, 381, 543

\bibitem[{{Wolf} \& {Sheinis}(2008)}]{wolf08}
{Wolf}, M.~J., \& {Sheinis}, A.~I. 2008, \aj, 136, 1587

\bibitem[{{Woo} {et~al.}(2008){Woo}, {Treu}, {Malkan}, \& {Blandford}}]{woo08}
{Woo}, J.-H., {Treu}, T., {Malkan}, M.~A., \& {Blandford}, R.~D. 2008, \apj,
  681, 925

\bibitem[{{York} {et~al.}(2000){York}, {Adelman}, {Anderson}, {Anderson},
  {Annis}, {Bahcall}, {Bakken}, {Barkhouser}, {Bastian}, {Berman}, {Boroski},
  {Bracker}, {Briegel}, {Briggs}, {Brinkmann}, {Brunner}, {Burles}, {Carey},
  {Carr}, {Castander}, {Chen}, {Colestock}, {Connolly}, {Crocker}, {Csabai},
  {Czarapata}, {Davis}, {Doi}, {Dombeck}, {Eisenstein}, {Ellman}, {Elms},
  {Evans}, {Fan}, {Federwitz}, {Fiscelli}, {Friedman}, {Frieman}, {Fukugita},
  {Gillespie}, {Gunn}, {Gurbani}, {de Haas}, {Haldeman}, {Harris}, {Hayes},
  {Heckman}, {Hennessy}, {Hindsley}, {Holm}, {Holmgren}, {Huang}, {Hull},
  {Husby}, {Ichikawa}, {Ichikawa}, {Ivezi{\'c}}, {Kent}, {Kim}, {Kinney},
  {Klaene}, {Kleinman}, {Kleinman}, {Knapp}, {Korienek}, {Kron}, {Kunszt},
  {Lamb}, {Lee}, {Leger}, {Limmongkol}, {Lindenmeyer}, {Long}, {Loomis},
  {Loveday}, {Lucinio}, {Lupton}, {MacKinnon}, {Mannery}, {Mantsch}, {Margon},
  {McGehee}, {McKay}, {Meiksin}, {Merelli}, {Monet}, {Munn}, {Narayanan},
  {Nash}, {Neilsen}, {Neswold}, {Newberg}, {Nichol}, {Nicinski}, {Nonino},
  {Okada}, {Okamura}, {Ostriker}, {Owen}, {Pauls}, {Peoples}, {Peterson},
  {Petravick}, {Pier}, {Pope}, {Pordes}, {Prosapio}, {Rechenmacher}, {Quinn},
  {Richards}, {Richmond}, {Rivetta}, {Rockosi}, {Ruthmansdorfer}, {Sandford},
  {Schlegel}, {Schneider}, {Sekiguchi}, {Sergey}, {Shimasaku}, {Siegmund},
  {Smee}, {Smith}, {Snedden}, {Stone}, {Stoughton}, {Strauss}, {Stubbs},
  {SubbaRao}, {Szalay}, {Szapudi}, {Szokoly}, {Thakar}, {Tremonti}, {Tucker},
  {Uomoto}, {Vanden Berk}, {Vogeley}, {Waddell}, {Wang}, {Watanabe},
  {Weinberg}, {Yanny}, \& {Yasuda}}]{york00}
{York}, D.~G., {et~al.} 2000, \aj, 120, 1579

\bibitem[{{Yu} \& {Tremaine}(2002)}]{yu02}
{Yu}, Q., \& {Tremaine}, S. 2002, \mnras, 335, 965

\bibitem[{{Zakamska} {et~al.}(2008){Zakamska}, {G{\'o}mez}, {Strauss}, \&
  {Krolik}}]{zakamska08}
{Zakamska}, N.~L., {G{\'o}mez}, L., {Strauss}, M.~A., \& {Krolik}, J.~H. 2008,
  \aj, 136, 1607

\bibitem[{{Zakamska} {et~al.}(2005){Zakamska}, {Schmidt}, {Smith}, {Strauss},
  {Krolik}, {Hall}, {Richards}, {Schneider}, {Brinkmann}, \&
  {Szokoly}}]{zakamska05}
{Zakamska}, N.~L., {et~al.} 2005, \aj, 129, 1212

\bibitem[{{Zakamska} {et~al.}(2003){Zakamska}, {Strauss}, {Krolik}, {Collinge},
  {Hall}, {Hao}, {Heckman}, {Ivezi{\'c}}, {Richards}, {Schlegel}, {Schneider},
  {Strateva}, {Vanden Berk}, {Anderson}, \& {Brinkmann}}]{zakamska03}
---. 2003, \aj, 126, 2125

\bibitem[{{Zakamska} {et~al.}(2006){Zakamska}, {Strauss}, {Krolik}, {Ridgway},
  {Schmidt}, {Smith}, {Heckman}, {Schneider}, {Hao}, \&
  {Brinkmann}}]{zakamska06}
---. 2006, \aj, 132, 1496

\end{thebibliography}


\end{document}